%% file: main.tex
\documentclass[reprint,superscriptaddress,amsmath,amssymb,aps]{revtex4-1}

\usepackage{graphicx} 
\usepackage{dcolumn}  
\usepackage{bm}       
\usepackage{layouts}  
\usepackage{braket}   
\usepackage{euscript} 
\usepackage{booktabs} 
\usepackage{hyperref} 

\usepackage{color}
\definecolor{deepblue}{rgb}{0,0,0.5}
\definecolor{deepred}{rgb}{0.6,0,0}
\definecolor{deepgreen}{rgb}{0,0.5,0}

\DeclareFixedFont{\ttb}{T1}{txtt}{bx}{n}{9} 
\DeclareFixedFont{\ttm}{T1}{txtt}{m}{n}{9}  

\usepackage{listings}
\newcommand\pythonstyle{\lstset{
language=Python,
basicstyle=\ttm,
commentstyle = \color{black},
otherkeywords={self},
keywordstyle=\ttb\color{deepblue},
emph={MyClass,__init__,circuit_searcher,circuit_reader},
emphstyle=\ttb\color{deepred},
frame=tb,
showstringspaces=false,
mathescape=true
}}
\lstnewenvironment{python}[1][]{\pythonstyle\lstset{#1}}{}

\newcommand{\name}{SCILLA }
\newcommand{\namecomma}{SCILLA, }
\newcommand{\namestop}{SCILLA. }

\renewcommand{\eqref}[1]{Eq.~(\ref{#1})} 
\newcommand{\figref}[1]{Fig.~\ref{#1}} 

\newcommand{\cel}[1]{\ensuremath{ \tilde{c}_{#1}^{-1}}}

\begin{document}

\title{Automated discovery of superconducting circuits and its application to\\ 4-local coupler design} 

\author{Tim Menke}
\email{tim\_menke@g.harvard.edu}
\affiliation{Department of Physics, Harvard University, Cambridge, MA 02138, USA}
\affiliation{Research Laboratory of Electronics, Massachusetts Institute of Technology, Cambridge, MA 02139, USA}
\affiliation{Department of Physics, Massachusetts Institute of Technology, Cambridge, MA 02139, USA}

\author{Florian H\"ase}
\affiliation{Department of Chemistry and Chemical Biology, Harvard University, Cambridge, Massachusetts, 02138, USA}
\affiliation{Department of Chemistry and Department of Computer Science, University of Toronto, Toronto, Ontario M5S 3H6, Canada}
\affiliation{Vector Institute for Artificial Intelligence, Toronto, Ontario M5S 1M1, Canada}

\author{Simon Gustavsson}
\affiliation{Research Laboratory of Electronics, Massachusetts Institute of Technology, Cambridge, MA 02139, USA}

\author{\\Andrew J. Kerman}
\email{ajkerman@ll.mit.edu}
\affiliation{MIT Lincoln Laboratory, 244 Wood Street, Lexington, MA 02420, USA}

\author{William D. Oliver}
\email{william.oliver@mit.edu}
\affiliation{Research Laboratory of Electronics, Massachusetts Institute of Technology, Cambridge, MA 02139, USA}
\affiliation{Department of Physics, Massachusetts Institute of Technology, Cambridge, MA 02139, USA}
\affiliation{MIT Lincoln Laboratory, 244 Wood Street, Lexington, MA 02420, USA}
\affiliation{Department of Electrical Engineering and Computer Science, Massachusetts Institute of Technology, Cambridge, MA 02139, USA}

\author{Al\'{a}n Aspuru-Guzik}
\email{alan@aspuru.com}
\affiliation{Department of Chemistry and Department of Computer Science, University of Toronto, Toronto, Ontario M5S 3H6, Canada}
\affiliation{Vector Institute for Artificial Intelligence, Toronto, Ontario M5S 1M1, Canada}
\affiliation{Canadian Institute for Advanced Research (CIFAR) Lebovic Fellow, Toronto, Ontario M5S 1M1, Canada}

\date{\today}

\begin{abstract}

Superconducting circuits have emerged as a promising platform to build quantum processors.
The challenge of designing a circuit is to compromise between realizing a set of performance metrics and reducing circuit complexity and noise sensitivity.
At the same time, one needs to explore a large design space, and computational approaches often yield long simulation times.
Here we automate the circuit design task using \namecomma a software for automated discovery of superconducting circuits.
\name performs a parallelized, closed-loop optimization to design circuit diagrams that match pre-defined properties such as spectral features and noise sensitivities.
We employ it to discover 4-local couplers for superconducting flux qubits and identify a circuit that outperforms an existing proposal with similar circuit structure in terms of coupling strength and noise resilience for experimentally accessible parameters.
This work demonstrates how automated discovery can facilitate the design of complex circuit architectures for quantum information processing.

\end{abstract}

\maketitle




\input{A-introduction}


\input{B-methodology.tex}


\input{C-results.tex}


\input{D-conclusion.tex}


\input{E-acknowledgements.tex}


\bibliography{refs}


\onecolumngrid
\clearpage

\input{F-appendix.tex}


\end{document}

%% file: A-introduction.tex
\section{Introduction}
\label{sec:introduction}

\begin{figure*}[htb]
	\centering
    \includegraphics[width = 1.0\textwidth]{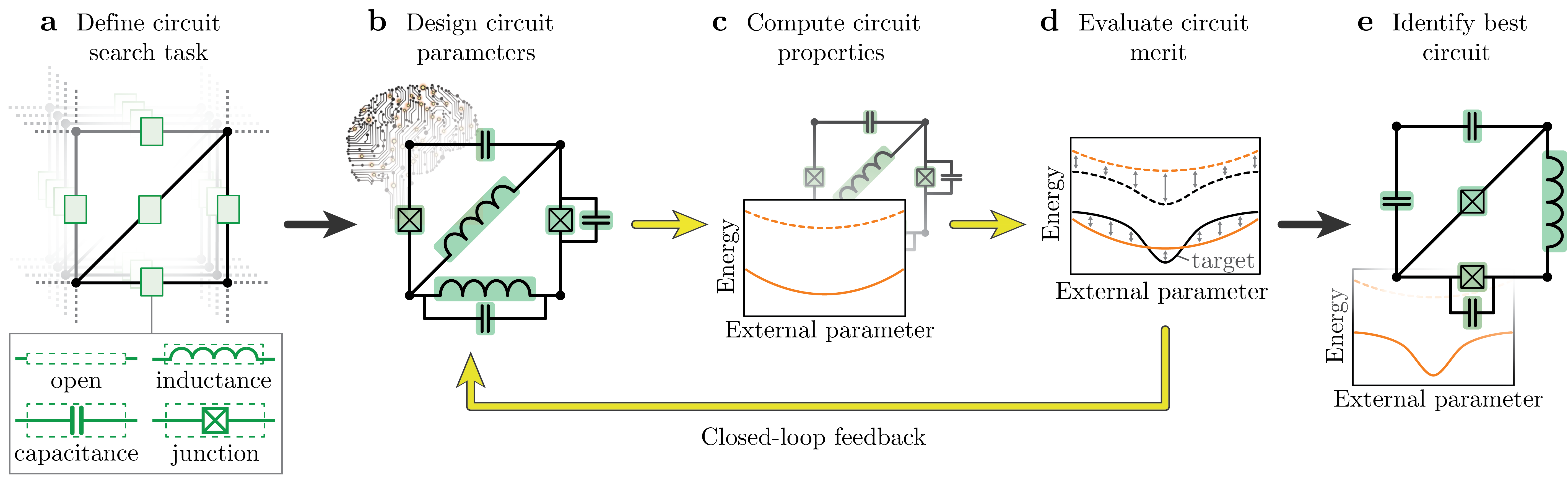}
    \caption{Implementation of \namecomma enabling automated circuit discovery. (a) Definition of the circuit design task, for which details about the general circuit architecture, parameter bounds, and design targets are provided. (b) Based on the general architecture, the design module evokes parameter generating algorithms to place components and choose component parameters in the circuit. (c) Calculation of circuit properties such as spectra or noise sensitivity. (d) Estimation the agreement between computed properties and target properties. (e) Circuits that are close to the design target are identified. A database system facilitates asynchronous execution and parallelization of the workflow as well as refinement of design choices (closed-loop feedback) based on merit evaluations of previously proposed circuits.}
    \label{fig:closed_loop_implementation}
\end{figure*}

The promise of quantum computing to surpass the capabilities of classical computers relies on a robust and scalable underlying hardware architecture.
In the context of quantum simulation and annealing, strong coupling, high connectivity, and many-body interactions between the qubits are necessary to accurately and efficiently represent the problem Hamiltonian \cite{cao2015hamiltonian, barkoutsos2017fermionic, babbush2014adiabatic}.
Superconducting circuits have proven to be a particularly well-suited platform due to their design versatility \cite{krantz2019quantum}.
Their quantum behavior arises from the interaction of modes that are set by effective inductances, capacitances, and non-linear Josephson junction elements in the circuit \cite{vool2017introduction}.
In this way, it is possible to design a wide variety of qubits and qubit-qubit coupling schemes at the circuit-diagram level and realize them in nanofabricated devices \cite{krantz2019quantum, oliver2013materials}.

A largely unexplored approach to meet circuit design challenges is to use computational automated discovery.
In other fields of science and engineering, automated discovery and inverse design have emerged as a solution to a variety of design problems.
All of these problems share the task to identify a set of parameters for which a system of interest yields desired target properties.
In contrast to forward design methods, where system properties are estimated from system parameters through direct measurements or computation, inferring the system parameters from target properties is a much more laborious process.
Typical automated discovery workflows consist of proposing a set of system parameters and determining the resulting system properties. 
The similarity between the obtained system properties and the targeted properties is then used as a quantitative measure to refine the system parameters iteratively.
Parameter refinements are usually implemented via various optimization procedures or reinforcement learning agents.
In the physical sciences context, automated discovery has been applied to nanophotonic on-chip devices \cite{lu2010inverse, piggott2015inverse}, complex quantum state generation in optical platforms \cite{krenn2016automated, melnikov2018active}, entanglement creation and removal in superconducting circuits \cite{goerz2017charting}, and further problems in many-body physics \cite{Chertkov:2018, Wigley:2016} and chemistry \cite{Gomez:2018, sumita2018hunting, zhavoronkov2019deep}.
This resulted in the discovery of conceptually new or previously unexplored solutions.

In the field of superconducting circuits, a contemporary design challenge is the implementation of many-body interactions.
Such interactions of more than two qubits commonly appear in effective spin models of quantum chemistry, quantum error mitigation schemes, and advanced driver Hamiltonians for quantum annealing \cite{babbush2014adiabatic, kandala2017hardware, bacon2006operator, fowler2012surface, hen2016quantum, martovnak2004quantum}.
Interactions involving $n$ qubits are called $n$-local and comprise Pauli operations in the system Hamiltonian that act on $n$ qubits simultaneously.
The interaction Hamiltonian has the form
\begin{equation*}
    H_\text{int} = M \prod_{i=1}^{n} \sigma_i^{\alpha_i},
\end{equation*}
where $2M$ is the coupling strength and $\alpha_i \in \{x,y,z\}$ denotes the type of Pauli matrix acting on each qubit.
Several implementations for 3- and 4-local couplers in superconducting circuits have been proposed.
Hamiltonian gadgets use ancilla qubits to encode the desired interaction in the effective low-energy spectrum of the system \cite{kempe2006complexity, chancellor2017circuit}.
Alternatively, individual flux qubits can be coupled to a common coupler circuit that mediates the interaction via a \textit{tailored potential} \cite{kerman2018mm, schondorf2018four, melanson2019tunable}.
The modularity of the tailored-potential approach in particular is reminiscent of existing 2-local couplers and holds promise for practical implementation \cite{Harris:2007, Weber:2017}.
Further work is required, however, to find coupler circuits that operate in experimentally accessible parameter regimes, reduce flux noise sensitivity, and eliminate spurious couplings.

The automated discovery task is to find a superconducting circuit diagram that fulfills a set of desired properties.
To this end, we introduce a method for \textbf{s}uperconducting \textbf{ci}rcuit c\textbf{l}osed-\textbf{l}oop \textbf{a}utomated discovery (SCILLA) and implement it in software.
We present the parallelized, closed-loop workflow that has access to design algorithms (generators) and can be connected to different circuit evaluators.
The method is applicable to a wide range of search problems involving spectral properties and noise sensitivities of galvanically connected superconducting circuits.
We demonstrate the performance and flexibility of \name on the well-studied example of capacitatively shunted flux qubits \cite{Yan:2016}.
We then highlight the competitive advantage of our approach by executing a discovery workflow which identifies noise-resilient 4-local coupling circuits.
As stated above, such a coupler would constitute a valuable building block for future quantum simulators and quantum annealers.

%% file: B-methodology.tex
\section{Methodology}
\label{sec:methodology}

The automated discovery workflow requires a circuit parameter generator, property evaluator, and merit estimator in order to fulfill a specified search task.
In the following, it is shown how circuit search can be formulated as an optimization problem by taking the circuit parameters as an input list and calculating the resulting closeness to the target properties.
We then describe the closed-loop implementation that handles parameter generation and optimization in a parallelized manner.
Lastly, 4-local coupler design is reduced to a spectral engineering problem that is compatible with the automated discovery workflow.

\footnotetext{A. J. Kerman, \textit{Efficient, hierarchical simulation of complex
Josephson quantum circuits,} in preparation.}
Throughout this work, the properties of superconducting quantum circuits are calculated by constructing and diagonalizing the circuit Hamiltonian.
For the automated discovery problem, and for spectral engineering problems more generally, the eigenenergies need to be calculated as a function of external degrees of freedom such as flux or charge offsets.
Circuit quantization has been discussed extensively in previous work, treating the circuit as a network of lumped elements with canonically conjugate flux and charge variables \cite{burkard2004multilevel, girvin2009circuit, vool2017introduction, krantz2019quantum}.
We automatically determine the Hamiltonian of a broad class of circuits and solve it efficiently by choosing a mixed representation in the charge and harmonic oscillator bases \cite{kerman2018mm, Note1}.

\subsection{Circuit design as an optimization problem}
\label{sec:optimizationproblem}

Automated discovery of circuits requires a quantitative metric to assess closeness of a candidate circuit to the target.
While the target can take a wide variety of forms, in this work we discuss and illustrate the practicality of optimizing specifically towards spectral properties, symmetry in the circuit network, and flux noise sensitivity.
It is straightforward to include more circuit properties as long as they can be obtained from Hamiltonian simulations in reasonable simulation time.
In addition, multiple target properties can be combined in a unifying merit function for more balanced design procedures.
We define the merit function as $\mathcal{M}:\,\mathbb{R}^d\rightarrow\mathbb{R}$, taking the circuit parameters as an input list $\mathbf{x}$ and returning a scalar value. 
It is defined such that a smaller function value corresponds to a circuit that better fulfills the target.
The specific functional form of $\mathcal{M}$ depends on the search task and needs to be carefully engineered for the optimization procedure to balance improvement of the sub-targets evenly.
It will be defined explicitly for each task in Sec.~\ref{sec:results}.
Since the target properties in general depend on the results of Hamiltonian circuit simulations, it is assumed that $\mathcal{M}$ has access to the simulation results for the input circuit.

The input $\mathbf{x}$ contains an ordered list of the capacitance, inductance, and junction energy for each circuit element between each pair of nodes.
If there is no circuit element between two nodes, the respective parameter value is zero.
Therefore, the ordered list of circuit element parameters fully defines the circuit network and is used to construct the circuit Hamiltonian.
The input also contains a list of external flux values -- one for each inductive loop in the circuit.
Each flux may only take the values $0.0\,\Phi_0$ or $0.5\,\Phi_0$, ensuring a symmetric spectrum as required for the applications presented in Sec.~\ref{sec:results}.
The circuit parameters are bounded, following typical experimentally accessible values \cite{Weber:2017, oliver2013materials}: up to $100\,\text{fF}$ for capacitances and up to $300\,\text{pH}$ for inductances.
The lower bound for inductances is chosen as $75\,\text{pH}$ in the 4-local coupler discovery in Sec.~\ref{sec:fourcoupler}.
This ensures that each inductor is large enough for qubits to couple mutually inductively to it.
For Josephson junction energies, we allow a range of $h\times 0-200\,\text{GHz}$ for the flux qubit benchmark in Sec.~\ref{sec:benchmark} and $h\times 99-1982\,\text{GHz}$ for the 4-local coupler discovery.
The latter provides more design flexibility and corresponds to junctions with $5\,\mu\text{A}/\mu\text{m}^2$ critical current density, $0.2\,\mu\text{m}$ junction width and $0.15-4.00\,\mu\text{m}$ junction length.
The intrinsic parallel capacitance of each Josephson junction depends on its respective physical geometry and is implicitly assumed.
Throughout this work, we denote the Josephson junction energy, capacitance, or inductance of a circuit element between two nodes $i$ and $j$ by $E_{\text{J}ij}$, $C_{ij}$, and $L_{ij}$, respectively.

In addition to the constraints on parameter values, we identify rules for the placement of components in the circuit network.
Capacitative elements may be placed in parallel to inductive elements such as junctions and inductors, i.e. between the same two nodes.
However, parallel placement of an inductance and a junction is not allowed to prevent the increased circuit complexity resulting from additional inductive loops formed in this way.
The degree of connectivity, i.e. the number of connections that each node has to other nodes, is set by the non-zero parameters in the input $\mathbf{x}$.
A circuit with more nodes and higher connectivity allows for more complex spectral engineering in reaching the target.
However, it is disadvantageous to increase the circuit size beyond the requirement of the objective.
Larger circuits may demand more calibration and more control hardware in experiments.
In addition, a larger number of external flux degrees of freedom usually leads to a greater flux noise susceptibility.
In the case of two- and three-node circuits -- excluding the ground node -- as discussed in this work, the circuit network is planar even for all-to-all connection between the nodes.
When moving to larger circuits, however, one needs to constrain certain circuit components to zero in order to maintain network planarity and thus ensure realizability in a planar on-chip architecture.
In principle, circuit complexity and experimental feasibility can also be added to the target with a suitable merit evaluation.

\subsection{Realizing closed-loop circuit discovery}

\name is a closed-loop implementation for accelerated computational circuit discovery available on GitHub \footnote{The automated discovery implementation for use with a generic circuit simulator (not provided) is available at the following address: \url{https://github.com/aspuru-guzik-group/scilla}}.
The procedure autonomously searches the circuit space defined in Sec.~\ref{sec:optimizationproblem} for circuit architectures satisfying desired target properties.
Similar approaches have already been successfully applied in the context of autonomous discovery and experimentation in chemistry and materials science \cite{Roch:2018a, Roch:2018b}.
\name contains a general-purpose method to compute properties of superconducting circuits \cite{Note1}.
It is thus applicable to a wide range of circuit design applications.

The typical workflow in the closed-loop implementation of \name is illustrated in Fig.~\ref{fig:closed_loop_implementation}.
Each circuit search is started from a set of general instructions, which inform \name about the size of the circuit and the number of components (see Fig.~\ref{fig:closed_loop_implementation}a).
With these specifications, a design algorithm chooses the kinds of components to be placed at particular locations in the circuit graph, as well as component-specific parameters (see Fig.~\ref{fig:closed_loop_implementation}b).
Then, properties of interest are computed for each designed circuit (see Fig.~\ref{fig:closed_loop_implementation}c).
As the calculation of circuit properties is typically the most time consuming step of this workflow, \name computes only the properties which are requested in the general instructions.
\name supports the computation of circuit spectra as well as estimations of the flux and charge noise sensitivites.
The computed circuit properties are then used to evaluate the merit $\mathcal{M}$, determining how well a given circuit matches the desired target specifications (see Fig.~\ref{fig:closed_loop_implementation}d).
Finally, the circuit composition which best satisfies the desired targets is determined (see Fig.~\ref{fig:closed_loop_implementation}e).
The crucial element to ``close the loop'' -- and thus enable autonomous circuit discovery -- is to report the merit of a proposed circuit architecture to the design algorithm.
Based on the feedback, the design algorithm can propose refined circuit architectures with improved target properties.
Some design tasks require higher-level operations, such as computing a gradient during circuit design or probing the flux sensitivity of a considered circuit only if a predefined condition is met.
Therefore, both the circuit design algorithms as well as the circuit merit evaluators are equipped with the capability to define new circuit evaluation tasks.
Circuit evaluations requested during circuit design are prioritized to assure that individual circuit design tasks are completed quickly.

The workflow presented in Fig.~\ref{fig:closed_loop_implementation} can be parallelized with little computational overhead by separating each step of the workflow into independent units (modules).
Information about individual circuits, such as component parameters or circuit properties, are dynamically stored in a system of databases built on the SQLite database engine.
The database implementation is the key component which allows the software to decouple each of the steps in the workflow and execute them individually. Details on the implementation are provided in the supplementary information (see Sec.~\ref{sec:closed_loop_implementation}).
The implicit parallelization of individual modules maximizes the number of circuits evaluated in parallel and thus utilizes available computing resources to higher capacity.
Moreover, circuit properties can be computed asynchronously, which is crucial as the computational time required to evaluate a given circuit often depends on its component configuration and parameter values.
With the implemented database system, all circuit parameters and properties which have been proposed and evaluated during the closed-loop sampling procedure are easily accessible.
As data is stored in a standardized format, profound post-process analysis of the circuits is possible and has the potential to identify general principles for circuit design, which will be demonstrated in the results (see Sec.~\ref{sec:results}). 

The design modules within \name can efficiently search the space of small circuits with relatively few components that can be evaluated quickly, as well as larger circuits with more components for which property calculations are more time consuming.
As the hardness of the optimization depends on the search space size and merit definition, a number of different design strategies are implemented in the closed loop, ranging from random search strategies \cite{Bergstra:2012,Bergstra:2011} for coarse, massively parallelizable sampling over gradient-based methods \cite{liu1989limited, Morales:2011} to evolutionary strategies \cite{Shi:1998, Eberhart:1995, miranda2018pyswarms}.
A more detailed description of the supported methods is provided in the supplementary information (see Sec.~\ref{sec:closed_loop_implementation}).

The closed-loop framework also enables the implementation of multi-step workflows.
The user declares a search procedure based on a chosen design algorithm, can define analyses on the circuits sampled during this first search, and then trigger another circuit design round, for instance, to refine the previous search.
Arbitrary combinations of different design and analysis steps are possible and can be easily integrated into the desired multi-step workflow.
Examples of such workflows are provided in the circuit discovery applications presented in Sec.~\ref{sec:results} as well as in the supplementary information (see Sec.~\ref{sec:multi_step_workflows}).

\subsection{4-local interaction as a spectral engineering problem}
\label{sec:4coupler_theory}

\begin{figure}
    \includegraphics[width = 1.0\columnwidth]{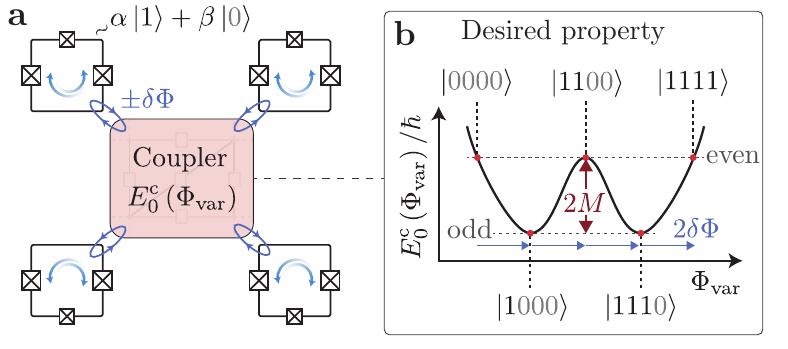}
    \caption{
    Coupler spectral property enabling 4-local interaction. (a) Four superconducting flux qubits are coupled mutually inductively to a shared coupler circuit. The states $\ket{0}$, $\ket{1}$ of each qubit are determined by the direction of the persistent circulating current in the qubit loop. By mutual induction, the magnetic field generated by the qubit's persistent current adds a small flux offset $\pm\delta\Phi$ to the coupler circuit. (b) The coupler has a ground state energy $E_0^\text{c}\left(\Phi_\text{var}\right)$ that depends on an external flux $\Phi_\text{var}$. If this energy spectrum follows a double-well shape with spacing $2\delta\Phi$ as indicated in the figure, the 4-qubit states with even and odd excitation numbers separate into two energy manifolds and the coupler mediates an effective 4-local interaction term. The challenge is to find a robust coupler circuit with such a double-well spectrum.
    }
    \label{fig:Intro2}
\end{figure}

The main application of \name presented in this work is the discovery of 4-local couplers for superconducting flux qubits.
We identify the key spectral property of such a coupler that yields the desired interaction when four flux qubits are coupled to its external flux degree of freedom via mutual inductance (see \figref{fig:Intro2}): a double-well profile of the coupler ground state energy $E_0^\text{c}$ versus external flux \cite{kerman2018mm}.
We note that this double-well flux spectrum of the coupler should not be confused with the inductive double-well potential of the flux qubit.
The ground state $\ket{0}$ and excited state $\ket{1}$ of each flux qubit are given by persistent left- and right-circulating currents in the qubit loop, respectively.
The circulating current change associated with the excitation (relaxation) of any of the flux qubits will shift the flux bias point $\Phi_\text{var}$ of the coupler by a small flux offset $+\delta\Phi$ ($-\delta\Phi$).
It is assumed that all four flux qubits are identical, have no pair-wise coupling, and have the same interaction strength with the coupler.
If the double-well profile of the coupler spectrum has flux spacings $2\delta\Phi$ between energy values differing by $2M$ as indicated in \figref{fig:Intro2}(b), then each qubit transition will raise or lower the potential energy of the system by $2M$.
Applying external flux such that $\Phi_\text{var}$ is biased at the center peak, the system energy separates into two energy manifolds associated with the parity of the qubit excitation number.
This is equivalent to the coupler providing a 4-local interaction term $H_\text{int} = M \, \sigma_1^z\sigma_2^z\sigma_3^z\sigma_4^z$.
The challenge of engineering a 4-local coupler therefore becomes that of finding a circuit with the described double-well spectrum in a narrow flux range under realistic parameter constraints.
In order to preserve quantum coherence, it is important to limit sensitivity to sources of noise, particularly flux noise if there are additional flux degrees of freedom in the coupler circuit \cite{gustavsson2011noise, krantz2019quantum}.
After a circuit fulfilling these properties has been found, simulations of the full system including all four qubits and the coupler need to be performed in order to confirm the validity of the 4-local coupling mechanism presented above.

%% file: C-results.tex
\section{Results \& Discussion}
\label{sec:results}

\begin{figure*}
    \centering
    \includegraphics[width = 1.0\textwidth]{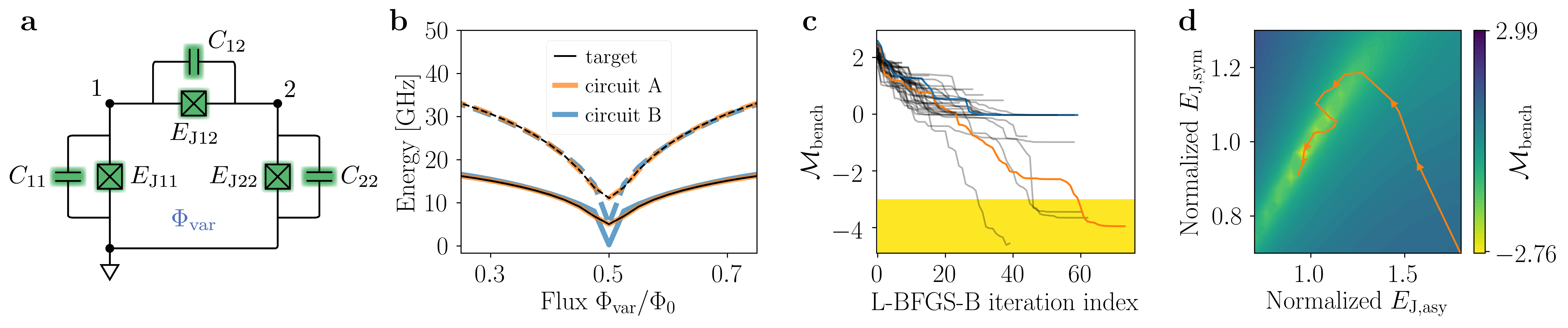}
    \caption{Flux qubit benchmark. (a) Parametrized circuit layout to which the circuit designer is constrained. (b) Target spectrum generated from a C-shunt flux qubit design, showing the transition energy from the ground state to the first (solid) and second (dashed) excited state. The final spectra of a successful (orange) and a stuck (blue) refinement run are also shown. (c) Convergence of 40 refinement runs starting at randomly sampled initial circuit parameters. (d) Visualization of a slice of the loss surface corresponding to the benchmarking problem. The orthogonally projected path of a successful refinement run towards the global minimum is shown.}
    \label{fig:Benchmark2}
\end{figure*}

We present the results of two circuit discovery instances.
First, \name is applied to a well-studied flux qubit example, testing the key features of the closed-loop implementation.
Knowledge of the optimal solution enables us to study the performance of the closed-loop algorithm in reaching the target.
Second, circuit discovery of 4-local couplers is performed.
The best circuit identified by the software is analyzed to elucidate its underlying operational mechanism, derive general design principles, and show its promise for experimental implementation.

\subsection{Benchmarking \name by flux qubit discovery}
\label{sec:benchmark}

As a benchmark for the automated discovery software, we define the target to be a capacitatively shunted (C-shunt) flux qubit.
This is a design variant of the flux qubit that has been shown to yield improved reproducibility and coherence for quantum information processing applications \cite{Orlando:1999, you2007low, Yan:2016}.
It is particularly well suited for strong qubit-qubit coupling in the quantum simulation and annealing context \cite{Weber:2017}.
The circuit is given by a loop of two identical, large junctions and one smaller junction, with the small junction shunted by a large capacitance.
It can be represented by a two-node circuit as shown in \figref{fig:Benchmark2}(a), with the additional bottom node declared ground.
The following Josephson junction energies and capacitances are typical for this type of qubit and are chosen as the target parameters:
\begin{align*}
    E_{\text{J}12}^\ast &= h\times 50\,\text{GHz} & E_{\text{J}11}^\ast &= E_{\text{J}22}^\ast = h\times 115\,\text{GHz}  \\
    C_{12}^\ast &= 45\,\text{fF} & C_{11}^\ast &= C_{22}^\ast = 0\,\text{fF}
\end{align*}
The intrinsic parallel capacitance of each junction is added to the listed capacitances.

The simulated transition energies between the ground state and the first -- and second -- excited states as a function of external flux through the circuit loop are shown in \figref{fig:Benchmark2}(b).
This energy spectrum is defined as the first target property for the circuit search problem.
There is, however, a continuous family of two-node circuits that fulfills this spectrum.
The degeneracy is lifted by adding the requirement that the circuit is symmetric, i.e. that the component parameter values $E_{\text{J}11}, C_{11}$ are close to $E_{\text{J}22}, C_{22}$, respectively.
Rather than enforcing a constraint a priori, circuit symmetry is added as a second target.
In this way, the design task tests a case of multi-objective search.
In addition, symmetry in the circuit diagram is a practically relevant property that can often be translated to the chip design, limiting noise-inducing effects such as currents in the ground plane.
It can be advantageous to have symmetry as a target property, rather than a constraint, such that it is only enforced when the other properties can be fulfilled at the same time.
A single scalar merit function $\mathcal{M}_\text{bench}$ is constructed from a combination of the spectrum sub-merit $\mathcal{M}_\text{bench}^\text{spec}$ and the symmetry sub-merit $\mathcal{M}_\text{bench}^\text{sym}$:
\begin{align*}
\mathcal{M}_\text{bench}^\text{spec} &= \sum_{i=1,2} \frac{\left|\left| E_{0i} - E_{0i}^\ast \right|\right|_{\mathcal{L}^2}^2}{2 \left(h\Phi_0 \cdot 1\,\text{GHz}\right)^2 } \\
\mathcal{M}_\text{bench}^\text{sym} &= \frac{\left| C_{11}- C_{22} \right|}{C^\text{max}} + \frac{\left| E_{\text{J}11} - E_{\text{J}22} \right|}{E_\text{J}^\text{max}} \\
\mathcal{M}_\text{bench} &= \log_{10}\left( \mathcal{M}_\text{bench}^\text{spec} + 100 \mathcal{M}_\text{bench}^\text{sym} \right)
\end{align*}
Here, $E_{0i}$ and $E_{0i}^\ast$ represent the candidate and target circuits' transition energy from ground to $i$th excited state as a function of external flux.
The RMS deviation $\left|\left|\cdot\right|\right|_{\mathcal{L}^2}$ between these functions is evaluated over the period of one flux quantum $\Phi_0$ and approximated numerically by computing the function values at discrete flux points.
The parameters $C^\text{max} = 100\,\text{fF}$ and $E_\text{J}^\text{max} = h\times 200\,\text{GHz}$ are the upper bounds for the respective parameter values as defined in Sec.~\ref{sec:optimizationproblem}.
The weighing of the sub-merits is chosen such that their values are on the same order for a typical random circuit, and the logarithmic scaling was observed to improve optimizer performance.
Therefore, the merit function is constructed by starting from a scalar form that allows for individual sub-merit optimization and then empirically choosing weighting and scaling parameters.

The workflow specified in the circuit searcher starts with random sampling from the parameter space, followed by gradient-based optimization with the L-BFGS-B algorithm \cite{liu1989limited, Morales:2011}.
Ten such search jobs are executed independently for greater throughput, each with 10 parallelized random samples and subsequent gradient-based refinement.
The parameter space for the benchmark is six-dimensional and consists of the junction energies and capacitances shown in \figref{fig:Benchmark2}.
No external flux needs to be specified, because the sole flux degree of freedom is varied to calculate the transition energy spectrum.
Moreover, the Hamiltonian of general two-node circuits without linear inductances takes a simple form than does not require the general Hamiltonian construction procedure included in \namestop
An optimized simulator for such two-node circuits is provided as a separate module and used here.
The simplified Hamiltonian is provided in Sec.~\ref{sec:supp:Htwonode} and a detailed implementation of the benchmark workflow is reported in Sec.~\ref{sec:multi_step_workflows}.

The result of 40 parallel circuit optimizations is shown in  \figref{fig:Benchmark2}(c), which represents a subset of four of the ten independent search jobs.
A fraction of $10\,\%$ of the runs (4 of 40) reach close to the global minimum defined by the target circuit.
High accuracy in both the spectral and the symmetry sub-targets is achieved in the yellow-shaded area in the plot, which is reached after 30-60 optimizer iterations for the successful runs.
The final spectrum of one successful run (circuit A) is shown in \figref{fig:Benchmark2}(b), matching the target spectrum accurately.
Symmetry in circuit A is very high, the deviation between the parameters $E_{\text{J}11}$ ($C_{11}$) and $E_{\text{J}22}$ ($C_{22}$) being $h\times 5\cdot10^{-9}\,\text{GHz}$ ($2\cdot10^{-8}\,\text{fF}$).

The remaining refinement runs terminate at a higher merit value.
Inspection of the final spectrum of one such run (circuit B) in \figref{fig:Benchmark2}(b) reveals that the spectrum is not close to the target.
We identify the failure mechanism in that the merit function has local minima.
In the surface projection of the merit function shown in \figref{fig:Benchmark2}(d), some local minima are visible as bright, yellow clusters around the global minimum.
Since the gradient-based optimizer used for refinement is local, it will naturally terminate after reaching one of the local minima.

In summary, nearly all module types and functionalities of \name have been tested in the benchmark.
In particular, it is verified that circuit simulations and merit evaluations are executed in parallel and asynchronously as intended.
The average run-time per search job is 6~h~41~min and evaluates $1.1\cdot 10^4$ circuits in total.
Therefore, when including the computational overhead of the closed-loop software, the average simulation time of a circuit is $2.3\;\text{s}/\text{circuit}$.
This compares well to $1.7\;\text{s}/\text{circuit}$ in a typical single, isolated evaluation of a circuit on the same hardware.
As a more general point about circuit discovery, we observe here that the optimization is non-convex even for a moderate search problem.
A purely gradient-based optimizer is of limited use in this case, although the random starting point ensures that the optimum can be found in some trials.
The success rate for finding good circuits is expected to decrease for larger circuits because of the increased parameter space dimensionality.
The hardness of the search problem is also increased by more restrictive merit functions with more sub-merits.
For the 4-local coupler search, the gradient-based optimizer is therefore replaced with an evolutionary strategy, which is able to avoid local minima.
In addition, the available computational resources and parallelization capability of \name are used to the maximum extent in order to explore a large portion of the design space and refine as many trial points as possible.

\subsection{Noise-insensitive 4-local coupler discovery}
\label{sec:fourcoupler}

\begin{figure}
\includegraphics[width = 1.0\columnwidth]{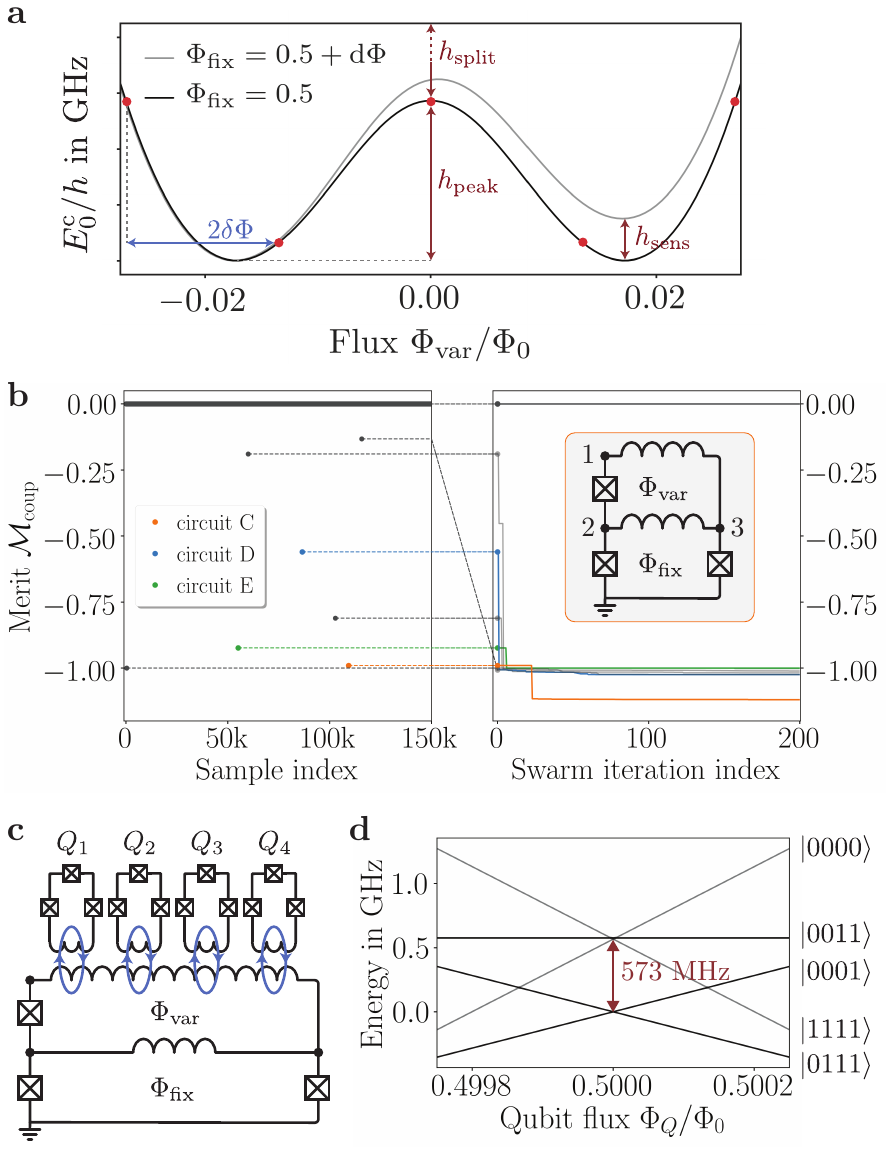}
\caption{
Discovery and verification of a 4-local coupler circuit. (a) Desired double-well spectrum, with the bias points for the 4-qubit excitation manifolds shown as red dots. The typical well asymmetry effect of flux offsets in the non-primary external fluxes is shown in gray. The merit function is constructed from the peak height $h_\text{peak}$, excited state splitting $h_\text{split}$, and noise sensitivity parameter $h_\text{sens}$. (b) Merit of the candidate circuits throughout the optimization, starting with 150,000 randomly sampled circuits and followed by refinement iterations using swarm optimization. Inset: circuit diagram of the best circuit (``circuit C") after refinement. (c) Circuit schematic of the full 4-qubit system including coupler circuit C. (d) First 16 energy eigenstates of the full system versus the commonly swept external flux in the qubit loops. Darker lines indicate a higher number of degenerate states. The states separate by qubit excitation parity at degeneracy, which corresponds to 4-local coupling between the qubits.
}
\label{fig:CouplerSearch}
\end{figure}

\footnotetext[5]{For a qubit-coupler mutual inductance $M$ and flux qubit persistent current $I_\text{p}^\text{qb}$, the equation $\delta\Phi = M I_\text{p}^\text{qb}$ is used to determine the flux offset induced by the qubit in the coupler.}
We turn to the design challenge of coupler circuits for 4-local interaction of flux qubits.
As detailed in Sec.~\ref{sec:4coupler_theory}, this problem can be reduced to finding a circuit with a double-well ground state spectrum.
The desired double-well profile is shown in \figref{fig:CouplerSearch}(a), with the bias points for the 4-qubit excitation manifolds shown in red.
The spacing $2\delta\Phi$ between the wells and the peak needs to be small such that it can be bridged by the spin flip -- and corresponding circulating current change -- of a mutually inductively coupled flux qubit.
Given the typical persistent currents and mutual inductances in strongly coupled flux qubit systems, we determine that the well-to-well spacing of the spectrum needs to be below $40\,\text{m}\Phi_0$ \cite{Weber:2017, Note5}.
In addition to the spectral property, insensitivity to noise sources in the system is required.
An often-observed effect of a flux offset in an additional inductive loop of the coupler circuit is shown as the gray trace in \figref{fig:CouplerSearch}(a):
The double-well spectrum tilts and becomes asymmetric, which would lead to unwanted, non-4-body terms in the interaction Hamiltonian.
Therefore, the second objective is to mitigate such spectral shifts due to flux noise.

In order to write these targets in a unified merit function, several parameters of the double-well are determined.
These parameters are visualized in \figref{fig:CouplerSearch}(a) (dark red).
In the computational routine, they are calculated in case a double-well feature is detected in a $40\,\text{m}\Phi_0$ range around the bias point of the primary external flux of the circuit:
First, the energy difference $h_\text{peak}$ between the wells and the center peak determines the 4-local coupling strength and needs to be maximized.
Second, the minimal energy difference $h_\text{split}$ between the ground and first excited state of the coupler is to be maximized; if it is too close to the qubits' transition energy,  the coupler can swap excitations with the qubits and break its operating principle of remaining in the ground state at all times.
Third, asymmetry induced by flux offsets in the non-primary loops of the coupler is calculated as the energy difference $h_\text{sens}$ between the minima of the left and right well.
Flux noise is usually dominated by local two-level systems on the metal surface of the circuit, with a degree of flux noise correlation between loops that depends on the length of wire shared by them \cite{koch2007fluxnoise, gustavsson2011noise, krantz2019quantum}.
An upper bound on the effect of flux noise is determined by applying flux offsets individually to each non-primary flux degree of freedom and summing the resulting asymmetries $h_{\text{sens},i}$.
A scalar merit function combines the above parameters.
It is given by the $p$-norm of three terms that quantify the peak height, excited state splitting, and noise insensitivity targets.
\begin{align*}
    \mathcal{M}_\text{coup} = - \left[ \left( \frac{\tilde{h}_\text{peak}}{h_\text{peak}^\text{max}} \right)^p + \left( \frac{\tilde{h}_\text{split}}{h_\text{split}^\text{max}} \right)^p + \left( 1 - \frac{\tilde{h}_\text{sens}}{\tilde{h}_\text{peak}} \right)^p \right]^\frac{1}{p}
\end{align*}
The hyperparameters $h_\text{peak}^\text{max}$, $h_\text{split}^\text{max}$ specify the target values for peak height and excited state splitting.
They serve as cut-off values for the respective parameters of the detected double-well as shown below.
The noise sensitivity is determined as the ratio between the summed asymmetries and the peak height, and can assume a maximum value of 1.
\begin{align*}
    \tilde{h}_\text{peak} &= \min\left\{ h_\text{peak}, \, h_\text{peak}^\text{max} \right\} \\
    \tilde{h}_\text{split} &= \min\left\{ h_\text{split}, \, h_\text{split}^\text{max} \right\} \\
    \tilde{h}_\text{sens} &= \min\left\{ \sum_\text{loops} h_{\text{sens},i}, \, \tilde{h}_\text{peak} \right\}
\end{align*}
The hyperparameters of the merit function are chosen empirically by trial of different hyperparameter sets.
For the results presented in this work, they assume the values $p = 4$, $h_\text{peak}^\text{max} = 1.5\,\text{GHz}$, and $h_\text{split}^\text{max} = 10\,\text{GHz}$.
The merit function is constructed such that a joint optimization of two sub-merits is favored over individual optimization.
The merit function assumes a minimum value of $-3^\frac{1}{p}$ if all sub-merits are maximally satisfied.
A merit of zero is assigned to the candidate circuit if no double-well is detected in the $40\,\text{m}\Phi_0$ flux range around the bias point, the peak height is below $50\,\text{MHz}$, or the excited state splitting is below $100\,\text{MHz}$.
In addition, a zero merit is assigned if the simulation fails, times out, or has large Hilbert space truncation errors.
Therefore, circuits with little promise for successful optimization are effectively removed from the optimization workflow.

The workflow implemented in the circuit searcher applies insights from the flux qubit benchmark to the more complex task of coupler design with flexible circuit diagrams.
It starts with random sampling of 15,000 circuits, which is about the limit beyond which database operations are observed to slow down.
The best two of the 15,000 circuits are kept after filtering and are refined using the evolution-inspired swarm optimization module.
Ten such jobs are executed independently for better utilization of the available computing resources.
The random search space is chosen to span three-node circuits with connections between the nodes and to ground on which capacitances, junctions, and inductances can be placed.
Under the constraints on placement of circuit components detailed in Sec.~\ref{sec:optimizationproblem}, components are randomly assigned among the available positions.
After sampling and filtering, the network configuration is fixed and only the component parameter values are refined.
The configuration of the circuit network is therefore flexible, and the number of inductive loops varies between sampled circuits.
This constitutes a consequential extension of the fixed-configuration, two-node search space that has been used for the benchmark, allowing for more degrees of freedom in reaching the 4-local coupler target.
To calculate the properties of such circuits, the simulation module that implements the general-purpose circuit Hamiltonian simulation is used.
The workflow implementation is described in more detail in Sec.~\ref{sec:multi_step_workflows}.

The results of the random search and subsequent swarm optimization are shown in \figref{fig:CouplerSearch}(b).
A double-well spectrum is detected in 7 of all 150,000 sampled circuits ($0.005\,\%$).
The swarm optimization improves the merit of all filtered circuits to varying degrees.
The final best circuit, named circuit C hereafter, has the following properties after (before) refinement:
\begin{align*}
    h_\text{peak}^\star &= 1.50\,\text{GHz} \, (1.24\,\text{GHz}) \\
    h_\text{split}^\star &= 0.87\,\text{GHz} \, (0.51\,\text{GHz}) \\
    h_\text{sens}^\star &= 0.20\,\text{GHz} \, (0.20\,\text{GHz})
\end{align*}
The circuit diagram is shown in the inset of \figref{fig:CouplerSearch}(b).
It is a two-loop circuit with the following non-zero component parameter values after refinement:
\begin{align*}
    C_{13}^\star &= 85.7\,\text{fF}  &  E_{\text{J}12}^\star &= h\times 1865\,\text{GHz}  &  L_{13}^\star &= 289\,\text{pH}\\
    C_{22}^\star &= 4.46\,\text{fF}  &  E_{\text{J}22}^\star &= h\times 196\,\text{GHz}  &  L_{23}^\star &= 120\,\text{pH}\\
    C_{23}^\star &= 16.8\,\text{fF}  &  E_{\text{J}33}^\star &= h\times 185\,\text{GHz}  & \\
    C_{33}^\star &= 70.6\,\text{fF} & & & 
\end{align*}

The double-well spectrum of circuit C emerges from the strong interaction of two rf SQUIDs at different flux bias points.
The first SQUID is formed by the loop surrounding the external flux $\Phi_\text{var}$, which is biased at $0\,\Phi_0$ in the absence of qubits.
The second rf SQUID loop encompasses the fixed external flux $\Phi_\text{fix} = 0.5\,\Phi_0$.
For details on the derivation of the circuit C Hamiltonian, see  supplementary Sec.~\ref{sec:supp:fourcoup_analytic}.
We note that this computer-generated circuit has strong similarities with the 4-local coupler design proposed by Kerman \cite{kerman2018mm}.
The key differences are that the parameters are more easily accessible experimentally in our design and the two inductive loops are connected galvanically -- and thus more strongly -- than for mutual inductive coupling, allowing for a large double-well peak despite smaller loop currents.
Moreover, the asymmetry relative to peak height arising from flux offsets is reduced by a factor of 3.3 in circuit C.

In total, 3 of the 7 double-well circuits from the automated discovery results have a similar layout as circuit C, with two inductive loops coupled through an inductance or junction.
The remaining 4 circuits are three-loop circuits and therefore have two non-primary external flux degrees of freedom.
In addition to circuit C, two more exemplary circuit optimization trajectories are highlighted in \figref{fig:CouplerSearch}(b): circuits D and E.
Circuit D is a three-loop circuit with 1.50\,GHz double-well peak and 3.12\,GHz excited state splitting.
These desirable properties are, however, contrasted by an increased flux noise sensitivity due to the additional external flux in the third loop.
Circuit E has two loops and a similar circuit network as circuit C but features a much larger excited state splitting of 10.8\,GHz and much smaller double-well peak height of 0.24\,GHz.
The properties of circuits C, D, and E are listed in detail in Sec.~\ref{sec:supp:fourcoup_spectra_diagrams}.
These examplary search results show that the automated discovery workflow finds a variety of circuits that fulfill the double-well sub-merits to different degrees.
It therefore supplies design options with different trade-offs, which can inform both theoretical understanding and practical implementation of the relevant class of circuits.

It remains to be demonstrated that circuit C in fact behaves as a 4-local coupler mediating an interaction $H_\text{int} = M \, \sigma_1^z\sigma_2^z\sigma_3^z\sigma_4^z$, and thus that our reduction of the design problem to a spectral property is valid.
For this reason, a full system simulation of four qubits coupled mutually inductively to the coupler as shown in \figref{fig:CouplerSearch}(c) is performed.
The energies of the 4-qubit excitation manifolds versus the qubit flux around degeneracy are shown in \figref{fig:CouplerSearch}(d), illustrating how the states of different parities separate.
A 4-local coupling strength of $2M = 573\,\text{GHz}$ in the circulating current basis ($\sigma_z$ basis) of the qubits is extracted, which is lower than the double-well peak height used as a proxy for the coupling.
Part of the reduction in coupling is caused by the flux points of the odd qubit excitation manifolds not lining up exactly with the minima of the double-well spectrum, as indicated conceptually by the red dots in \figref{fig:CouplerSearch}(a).
Further reduction mechanisms could include inductive loading of the coupler by the mutual inductive coupling to the qubits and interactions with the coupler's excited state.
Spurious terms of different locality, which are manifest as a splitting of states at the additional spectral crossing points in \figref{fig:CouplerSearch}(d), are in the MHz-regime and thus negligible.
We conclude that despite additional effects reducing the coupling, \name applied to the double-well merit was able to design a coupler with several hundred MHz of 4-local coupling strength, even without performing costly full system simulations.

As predicted from the benchmark analysis, our success in finding a promising 4-coupler circuit rests on careful definition the merit function, choice of the discovery workflow, and exploitation of computational resources.
Given the low success rate of sampling a circuit with double-well spectrum at all, a critical step has been to explore a large portion of the design space before attempting to refine promising circuits.
Averaged over the ten batches, sampling 15,000 circuits took $21.2\pm 1\,\text{h}$ ($5.09\,\text{s/circuit}$).
The subsequent swarm optimization with a total of 8,000 circuit evaluations in 200 iteration steps took an average of $51.2\pm 15.9\,\text{h}$ ($23.0\,\text{s/circuit}$).
While the sampling time is relatively uniform, the swarm optimization runtime varies significantly with the circuit layout that is being optimized.
In addition, it is observed that the simulation time per circuit is much longer in the swarm optimization, which is expected from the lower degree of parallelization, additional circuit simulations required by the merit evaluation, and larger number of database operations in the iterative optimization.
On top of the runtimes of the sampling and optimization steps, the closed-loop procedure requires an additional $3.30\pm 0.37\,\text{h}$ that is mainly spent filtering the sampled circuits before swarm optimization.
The filtering time is limited by the input/output operation speed of the used hardware and can be much lower on a different computing cluster.

Overall, \name is able to accommodate the increased computational effort from adding a third node to the circuit and allowing for full flexibility for the circuit network.
The added complexity is rewarded by successful fulfillment of the target properties.
This warrants further study of improving the efficiencies of circuit simulation and search algorithms.

%% file: D-conclusion.tex
\section{Conclusion}

We demonstrated a computer-driven and highly parallelized approach to the design of superconducting circuits.
We evaluated the performance of the developed method \name and discovered circuits for design challenges with multiple objectives.
Our results demonstrate that \name is successful in finding new circuit architectures with superior performance than an existing proposal, namely a 4-local coupler with several hundred MHz of coupling strength, small unwanted coupling terms, small flux noise sensitivity, and experimentally accessible parameters.
Given the modularity of the software implementation, it is straightforward to tackle new circuit design challenges that can be defined in terms of spectral properties and eigenstate expectation values.
The exponential scaling of the Hilbert space with the number of nodes makes na\"{i}ve scaling of the method by simply adding more nodes to the circuit impractical.
However, simulation of larger systems is possible if they consist of small sub-circuits with inter-circuit coupling that is weaker than the intra-circuit coupling \cite{kerman2018mm, Note1}.
In that case, the sub-circuit Hamiltonians may be solved individually with only the lowest eigenstates of the sub-circuits then coupled together, enabling inverse design of other common architectures such as transmon-based multi-qubit processors.
In addition, inspired by Krenn \textit{et al.} \cite{krenn2016automated}, one can envision functionality to store discovered circuits as building blocks for larger architectures, thus enlarging the set of available circuit components.
Future work also entails translation of the most promising architectures into actual chip designs in order to validate the targeted properties experimentally.
This will allow us to feed back on the computational routine and include additional practical constraints in the circuit search.

%% file: E-acknowledgements.tex
\section{Acknowledgements}

We acknowledge J. Braum\"uller, A. Di Paolo, C.F. Hirjibehedin, M. Krenn, T.P. Orlando, S. Sim, S.J. Weber, and R. Winik for valuable discussions.
This research was funded in part by the Office of the Director of National Intelligence (ODNI), Intelligence Advanced Research Projects Activity (IARPA), via U.S. Army Research Office Contract No. W911NF-17-C-0050; and by the Department of Defense via MIT Lincoln Laboratory under U.S. Air Force Contract No. FA8721-05-C-0002.
F.H. was supported by the Herchel Smith Graduate Fellowship and the Jacques-Emile Dubois Student Dissertation Fellowship.
A.A.-G. acknowledges the generous support from Google, Inc. in the form of a Google Focused Award.
T.M., F.H. and A.A.-G. acknowledge financial support from the Canada 150 Research Chairs Program as well as Dr. Anders Fr{\o}seth.
All computations reported in this paper were completed on the Odyssey cluster supported by the FAS Division of Science, Research Computing Group at Harvard University.
The views and conclusions contained herein are those of the authors and should not be interpreted as necessarily representing the official policies or endorsements, either expressed or implied, of the ODNI, IARPA, or the U.S. Government. The U.S. Government is authorized to reproduce and distribute reprints for Governmental purposes notwithstanding any copyright annotation thereon.

%% file: F-appendix.tex
\section*{Supplementary Information}
\appendix 
\renewcommand\thesection{S.\arabic{section}} 
\renewcommand\thesubsection{S.\arabic{section}.\arabic{subsection}}
\renewcommand\thefigure{S.\arabic{figure}}
\renewcommand\thetable{S.\arabic{table}}  


\section{Implementation details of the closed-loop approach to circuit discovery}
\label{sec:closed_loop_implementation}
    
\name constitutes a flexible approach to the automated discovery of superconducting circuits.
The implementation relies on an abstraction of the circuit design process to enable parallelizable circuit design, where different design strategies and evaluation methods can be combined in multi-step workflows for accelerated discovery.
The abstraction level implemented in \name separates the design process into three major steps.
First, circuit compositions need to be suggested, as well as parameters for individual components.
Then, circuit properties are computed. Finally, based on the computed properties, the merit of the circuit is evaluated.
This separation enables the definition of computational modules which execute each of the steps individually and independently from other modules.
While every circuit will pass through every single step in the closed-loop, circuits can be scheduled for different steps depending on the availability of the modules, thus enabling asynchronous parallelization of the circuit design process.
Note that modules are equipped with the ability to request new circuit evaluation tasks, e.g. to determine a gradient or to quantify the noise sensitivity of a designed circuit.
Circuit evaluations requested by modules of \name are always prioritized over circuit design tasks requested by the researcher to assure that individual circuit design tasks are completed quickly. 
  
The closed-loop approach to circuit discovery consists of three major modules that mirror the steps described above:
(i) the design module, which suggests new circuit compositions and component parameters, (ii) the property calculation module, which computes properties for a circuit with a given set of components and external fluxes, and (iii) the merit evaluation module, which assesses the agreement of the computed properties with the target properties quantitatively. All modules provide interfaces to a number of methods for their specific purposes, which are described in more detail in this section.

\subsection{Design module}\label{sec:design_module}

The design module suggests components and component parameters for circuit architectures.
For this task, the design module requires information about the number of components which can be placed at different locations in the circuit graph as well as the accepted ranges for parameter values of these components.
The closed-loop implementation enables the design module to access the estimated merit of circuit architectures it designed.
In this way, it is able to refine circuit architectures and make more qualified decisions about which circuits to probe next. 
The design module supports a number of different design algorithms for different application scenarios.
The application scenarios we considered include cases where circuit properties can be computed rapidly as well as cases where the computation of circuit properties is computationally expensive.
In addition, we distinguish cases where precise quantitative agreement with the desired target properties is required and cases where differences between the targets and the achieved properties are tolerated to some degree. Combinations of these cases can be considered by implementing multi-step circuit search workflows (see Sec.~\ref{sec:multi_step_workflows}). In the following sections, we provide a detailed list of supported design algorithms along with scenarios to which we believe the design algorithms are most applicable.\\

\paragraph{Random search optimization\\} 

Random search optimization follows the na\"ive strategy of proposing new parameters by drawing parameter points from a distribution which is supported on the entire parameter space.
Without any prior assumptions about the location of the global optimum, parameter points are typically sampled from the uniform distribution as the optimum could be located anywhere equally likely.
Aside from not making any assumptions about the shape of the response surface, random search optimization features two additional important characteristics:
(i) Generating random numbers for standard probability distributions such as the uniform distribution is computationally cheap, and (ii) the decision about which new parameter set to propose does not depend on prior evaluations.
As such, random search is massively parallelizable and particularly suited for cases where circuits consist of only a few components and properties can be computed quickly.
Random search can therefore be used to rapidly screen the parameter space without a bias towards the exploitation of prior evaluations.
As shown in the main text, it is also suitable to pre-screen large parameter spaces with slow evaluation times for the purpose of finding a few initial guesses that are further refined subsequently.\\

\paragraph{Gradient-based optimization\\}

Gradient-based optimization approaches aim to locate the optimum of an objective function using information about the gradient (and possibly higher order derivatives) of the function at particular positions. Starting from an initial parameter point in the parameter domain, gradient-based optimization algorithms evaluate the gradient of the objective function at this point to guess the search direction and the step distance to the optimum, which is then proposed as a new point in parameter space. 

In contrast to random search algorithms, gradient-based algorithms account for a number of prior evaluations when making decisions about which parameter point to evaluate next.
Gradient-based algorithms are therefore capable of exploiting previously seen evaluations to some extent, which is of use particularly when good circuits are hard to find and slow to evaluate.
However, as the decision-making process relies on information acquired from the local environment of the current parameter point, gradient-based optimization methods typically only locate minima in the convex environment of the starting point.
If the objective landscape is non-convex, gradient-based optimization algorithms are prone to get stuck in a local optimum instead of locating the global optimum.
Nevertheless, gradient-based algorithms can locate local optima to a high degree of accuracy.
Based on this feature, we recommend to use gradient-based optimization for fine-tuning circuit architectures predicted by less accurate design algorithms such as random search, i.e. to refine the component parameters.
However, this method is expected to succeed only for search problems with few parameters, as high-dimensional circuit search spaces often have an abundance of local minima.

\name supports optimizations with the Broyden-Fletcher-Glodfarb-Shanno (L-BFGS-B) algorithm [S1].
This algorithm is a quasi-Newton method that approximates the Hessian based on results from a number of gradient evaluations.
It is thus guaranteed to converge for functions which feature a quadratic Tailor expansion around the global optimum, but has generally shown good performance even for non-smooth optimizations [S2].\\

\paragraph{Nature-inspired optimization\\}

Nature-inspired optimization algorithms are based on various phenomena encountered in nature, most notably biology and physics. As such, nature-inspired optimization algorithms can differ greatly in their mechanics, and include evolutionary strategies as well as swarm intelligence methods or annealing approaches.
One prominent example is particle swarms optimization [S3,\,S4]. This optimization strategy relies on a population of candidate solutions (particles), which are iteratively modified based on the merit of each particle.
The modification of an individual particle commonly accounts for the particle's individual position, momentum, and merit, and the position and merit of the most promising other particle.
Particle swarm methods typically contain a number of hyperparameters which determine the extent to which each of the aforementioned aspects contribute to the overall update of a particle's position.
The particular choice of hyperparameters can greatly affect the optimization performance, and particle swarm methods are not guaranteed to locate the global optimum, despite performing well empirically.

\name supports the particle swarms implementation PySwarms by Miranda [S5], which is shown in this work to perform well even in the large parameter spaces encountered in 4-local coupler discovery.
Unlike in the L-BFGS-B implementation, where circuit evaluations are performed iteratively to find the gradient, the gradient-free swarm optimization enables parallelization within each iteration step.
Therefore, it is possible to re-use computational resources from earlier steps in a multi-step workflow by adding additional particles.
This constitutes another example of how the multi-step workflow capability of \name can leverage the advantages of several optimization strategies for accelerated circuit discovery.

We observe that PySwarms evaluates some parameter sets repeatedly within one optimization procedure, especially when parameter values are close to the bounds of the parameter range.
In order to increase computational efficiency, returning to the same parameter value should be avoided.
We leave this software improvement to future work.

\subsection{Property computation module}

The property computation module operates on circuits for which the layout and component parameters have already been proposed.
Based on the circuit architecture, the eigenenergies are computed for a range of specified external fluxes.
Functionality can be added to compute operator expectation values, although this is not required for the search tasks considered in this work.
The procedure for determining and diagonalizing general flux-based circuit Hamiltonians is referenced in Sec.~\ref{sec:methodology} in the main text.
The simplified Hamiltonian for two-node circuits that is used for the flux qubit benchmark is documented in Sec.~\ref{sec:supp:Htwonode}.
An implementation of the two-node Hamiltonian simulation is included in the \name GitHub repository [S6].

As circuit property calculations can be computationally costly and eventually become the bottleneck of the design process for sufficiently large circuits, the property computation module encapsulates the property computation operation into a stand-alone process. This process can be executed locally, independently from the closed-loop, or potentially even sent to a distributed computing platform such as a separate computing cluster or a cloud based solution.

\subsection{Merit evaluation module}

The merit evaluation module estimates how well the properties of a proposed circuit architecture align with the target properties.
As such, the merit evaluator determines the response landscape on which the design algorithms operate.
The scalar merit functions for the flux qubit benchmark and 4-local coupler search are presented in Sec.~\ref{sec:benchmark} and Sec.~\ref{sec:fourcoupler} in the main text, respectively.
The merit evaluation module can request new tasks, which is used to calculate the noise sensitivity in the 4-local coupler discovery.
In this case, the circuit is again simulated for a small change in one of the non-primary external fluxes.
As described in Sec.~\ref{sec:closed_loop_implementation}, circuit evaluations requested by a module such as the merit evaluation module are prioritized over new circuits requested by the researcher.

\subsection{Closing the loop}

Asynchronously parallelized execution of the circuit design process is implemented via a separate module supervising the design, property and merit modules.
This master module schedules individual circuits for each of the steps in the automated discovery process depending on the availability of the three modules.
The scheduler aims to maximize the load on each individual module to maximize the throughput in the design process.
Information about individual circuits such as composition, parameter values, or properties are synchronized with a database framework built on the SQLite database system at every step.
The database framework consists of a number of individual databases, which contain only the information relevant to individual modules.
This allows for further parallelization of the design workflow, as different modules can access different databases at the same time.
In addition, caching reduces the overhead due to input/output operations. 
  

\section{Multi-step workflows}
\label{sec:multi_step_workflows}

\name supports a variety of different optimization strategies for designing superconducting circuits (see Sec.~\ref{sec:closed_loop_implementation}.1). Multi-step workflows -- in which a number of different optimization strategies is used synergistically for the same design task -- divide the design task into sub-problems in which the individual strengths of different optimization algorithms can be leveraged. \name provides an API for the simplified deployment of such multi-step workflows. 

\subsection{C-shunt flux qubit example}
\label{sec:supp:benchmark_workflow}
    
In this section, we describe how the multi-step workflow for the C-shunt flux qubit benchmark in Sec.~\ref{sec:benchmark} is constructed.
This example combines a coarse scan of the parameter space with a subsequent refinement of the circuits obtained from the coarse scan.
The coarse scan is implemented as a random search and the L-BFGS-B optimization algorithm is used to refine the architectures.
    
The definition of this multi-step workflow starts with the declaration of a \texttt{CircuitSearcher} object. This object provides the application programming interface (API) for \name and its functionalities. To initialize a \texttt{CircuitSearcher} object, we first need to provide some settings regarding the general layout and parameter bounds for circuits considered in this circuit search (see Fig.~\ref{fig:circuit_param_declaration}). \\
    
\begin{figure}[!ht]
\begin{python}
# Declare circuit layout and parameters bounds
c_specs        = {`dimension': 3, `low': 0., `high': 100., `keep_prob': 1.}
j_specs        = {`dimension': 3, `low': 0., `high': 200., `keep_num':  3}
circuit_params = {`c_specs': c_specs, `j_specs': j_specs, `l_specs': None,
                  `phiOffs_specs': None}
\end{python}
\caption{Declaration of general layout and parameter bounds. Parameters for capacitances, inductances and junctions are declared as dictionaries. The \texttt{dimension} key indicates how many instances of each compound type can maximally be placed in the circuit. The \texttt{low} (\texttt{high}) key provides a lower (upper) bound on the parameter values. The \texttt{keep\_prob} and \texttt{keep\_num} keys can be used to increase the sparsity of the circuit layout: \texttt{keep\_prob} specifies the probability that a circuit element of the respective type will be placed between two nodes, and \texttt{keep\_num} is the total number of circuit elements of this type to be placed in the circuit. Capacitances are declared in fF, inductances in pH, and junction energies in GHz. The \texttt{phiOffs\_specs} key, which is not used in the flux qubit example, specifies possible values for external fluxes and is only relevant for the general circuit simulator.}
\label{fig:circuit_param_declaration}
\end{figure}
    
In addition to circuit component parameters, additional general parameters are declared. Those provide information about how the closed-loop implementation should be executed and which approximations can be made. For this example, we choose to employ the two-node circuit simulator and specify the external flux to sweep a full flux quantum. \\
    
\begin{figure}[!ht]
\begin{python}
# Declare circuit searcher settings
general_params = {`solver': `2-node', `phiExt': np.linspace(0, 1, num = 41)}
\end{python}
\caption{Declaration of general parameters for the circuit search. General parameters are declared as a dictionary, in the same way as circuit parameters (see Fig.~\ref{fig:circuit_param_declaration}).}
\label{fig:general_param_declaration}
\end{figure}
    
With these two parameter dictionaries, an instance of the \texttt{CircuitSearcher} is initialized (see Fig.~\ref{fig:circuit_searcher_initialization}).
Importantly, the circuit searcher instance has two different use cases:
It can be used to (i) implement a single- (or multi-) step workflow, or (ii) connect to an existing database to analyze results of a previously executed workflow.
Here we first discuss workflow implementation and execution.\\
    
\begin{figure}[!ht]
\begin{python}
# Initialize circuit searcher
circuit_searcher = CircuitSearcher(circuit_params, general_params,
                                   database_path = `Experiments')
\end{python}
\caption{Initialization of the circuit searcher. The circuit searcher provides an interface to \name and its functionalities. The path to the database folder is specified with the option \texttt{database\_path}.}
\label{fig:circuit_searcher_initialization}
\end{figure}

We can now add the first step of our multi-step workflow to the circuit searcher instance. Our goal for the first step is to perform coarse sampling of the parameter space using the random designer. We will query a total of 10 randomly generated circuit architectures.
For each constructed circuit, we compute the spectrum and compare it to a desired target spectrum.
The \texttt{add\_task} method of the circuit searcher is used to queue the task to the internal task list of the circuit searcher (see Fig.~\ref{fig:add_random_search_task}).
This does not yet execute the random search task.\\

\begin{figure}[!ht]
\begin{python}
# Specify settings for the random designer
designer_options = {`max_iters': 10, `max_concurrent': 10, `batch_size': 10}
    
# Specify merit function
target_spectrum  = ... <defined elsewhere> ...
merit_options    = {`target_spectrum': target_spectrum, `include_symmetry': True}
    
# Add task to the task queue
computing_task_1 = $\color{deepred}\text{\ttb circuit\_searcher}$.add_task(name = `random_search', 
                         designer = `random',         designer_options = designer_options, 
                         merit    = `TargetSpectrum', merit_options    = merit_options)
\end{python}
\caption{Adding a random search task to the circuit searcher, with the goal to design circuit architectures matching a desired target spectrum. We first set options for the design task: We specify that the design procedure is run for 10 iterations, computing properties for at most 10 circuits at a time. Every time the random designer is queried, it will generate 10 individual circuits. These are tested for validity by checking invertibility of the capacitance matrix before being passed on to the property computation module. Circuits are evaluated based on their spectrum, taking into account the symmetry of the circuit configuration. The merit function specified in \texttt{merit} is defined in \textsl{critic\_target\_spectrum.py} in the folder \textsl{CircuitQuantifier} (see GitHub repository [S6]).}
\label{fig:add_random_search_task}
\end{figure}

In the next step, we define a filtering task (see Fig.~\ref{fig:filtering_task}).
This task accesses the database and queries all merits for all stored circuits.
Circuits are then labeled based on their merit for subsequent operations.
Typically, one would determine the best-performing circuit architectures of the previous design step and pass only those on to the next step in the workflow.
In this particular example, however, we decided to keep all sampled circuits such that the benchmark evaluates performance of the optimization for a larger variety of starting points.\\

\begin{figure}[!ht]
\begin{python}
# Filter the randomly sampled circuits
filtering_task_1 = circuit_searcher.add_task(name = `filtering', designer = `filter_db',
                                             designer_options = {`num_circuits': 10})
\end{python}
\caption{Declaration of a filtering task to determine the best-performing circuits obtained from previous design tasks. In the flux qubit example, all 10 circuits from the random search step are kept.}
\label{fig:filtering_task}
\end{figure}

In the second design step, the L-BFGS-B optimization algorithm is employed (see Fig.~\ref{fig:lbfgs_design}).
This design step is executed for a total of 100 iterations of the L-BFGS-B algorithm.
Again, we aim to discover circuit architectures that match the target spectrum.
In contrast to the random design step, we now set the \texttt{use\_library} flag of the \texttt{add\_task} method to indicate that this design step starts from the circuits which were identified in the previous filtering task.
As all sampled circuits were kept in the filtering, this is not necessary here but will be in most other scenarios.\\

\begin{figure}[!ht]
\begin{python}
# Specify settings for the L-BFGS-B designer
designer_options = {`max_iters': 100, `max_concurrent': 10}
    
# Specify merit function
target_spectrum  = ... <defined elsewhere> ...
merit_options    = {`target_spectrum': target_spectrum, `include_symmetry': True}
    
# Add task to the task queue
computing_task_2 = circuit_searcher.add_task(name = `lbfgs_optimization', 
                         designer = `scipy',          designer_options = designer_options, 
                         merit    = `TargetSpectrum', merit_options    = merit_options,
                         use_library = True)
\end{python}
\caption{Implementation of the L-BFGS-B refinement design step. The declaration of this design step closely follows the declaration of the random design step (see Fig.~\ref{fig:add_random_search_task}). However, we use the \texttt{use\_library} flag to indicate that this design step should be started based on the circuits identified in the previous filtering task.}
\label{fig:lbfgs_design}
\end{figure}

All steps that were defined in the multi-step workflow can be executed sequentially by calling the \texttt{execute} method of the circuit searcher (see Fig.~\ref{fig:execute_circuit_searcher}). For each individual step, this method schedules tasks for the modules (design, property computation, and merit evaluation), synchronizes the execution of the workflow, and stores intermediate and final results in the databases for later access and analysis. \\

\begin{figure}[!ht]
\begin{python}
# Run the specified multi-step workflow
circuit_searcher.execute()
\end{python}
\caption{Running the \texttt{execute} method of the circuit searcher executes all previously defined steps.}
\label{fig:execute_circuit_searcher}
\end{figure}

After execution of the workflow, the results can be read on any machine as long as the database is available.
Importantly, database readout can be done on a separate thread.
Towards this end, a new \texttt{CircuitSearcher} object is initialized (see \figref{fig:initialize_circuit_reader}).
As no new circuit search job is initialized, only the database path needs to be specified.

\begin{figure}[!ht]
\begin{python}
# Initialize CircuitSearcher object for database readout
circuit_reader = CircuitSearcher(database_path = `Experiments')
\end{python}
\caption{Initialization of a \texttt{CircuitSearcher} object for readout of a database stored in \textsl{Experiments}, which is a sub-directory of the current working directory.}
\label{fig:initialize_circuit_reader}
\end{figure}

Subsequently, the circuits that were generated in the sampling and L-BFGS-B optimization steps are read from the database (see \figref{fig:read_database}).
Using the \texttt{query} method, we first extract the unique identifiers that were assigned to the design steps at runtime.
Provided with the step identifier, the same method then reads a list of all sampled circuits and a list of L-BFGS-B optimization trajectories.
The trajectories are stored in a dictionary that contains an identifier and an ordered list of circuits that were encountered in the optimization for each optimization run.

\begin{figure}[!ht]
\begin{python}
# Read list of design steps
computing_tasks  = circuit_reader.query(kind = `list_computing_tasks')

# Read sampled circuits
sampled_circuits = $\color{deepred}\text{\ttb circuit\_reader}$.query(kind = `get_circuits_from_task', 
                                        task = computing_tasks[1])

# Read LBFGS-B optimization trajectories
lbfgs_trajectories = circuit_reader.query(kind = `get_trajectories',
                                          task = computing_tasks[2])
\end{python}
\caption{Readout of the database. Depending on the \texttt{kind} key provided, the \texttt{query} method can read the design step identifiers, list of sampled circuits, or dictionary of L-BFGS-B optimization trajectories.}
\label{fig:read_database}
\end{figure}

Each circuit in the extracted lists is represented as a dictionary that contains the following information: circuit layout and parameters, computed circuit properties such as the spectrum, merit value, other circuits that were evaluated to compute the merit, as well as various unique identifiers that were used for bookkeeping and database handling during the closed-loop computation.
This allows for post-processing and further analysis of the discovered circuits.

\subsection{4-local coupler example}

In this section, we detail the automated discovery settings for the 4-local coupler discovery discussed in Sec.~\ref{sec:fourcoupler}.
The same \texttt{CircuitSearcher} methods are used as in the flux qubit example in Sec.~\ref{sec:multi_step_workflows}.1, we just need to update the settings and parameters.
As shown in \figref{fig:initialization_fourcoupler}, a \texttt{CircuitSearcher} object is initialized for a three-node circuit search with parameter bounds as specified in Sec.~\ref{sec:optimizationproblem}.
In addition, the general-purpose circuit solver (named ``JJcircuitSim") that is referenced in Sec.~\ref{sec:methodology} in the main text is specified for circuit spectrum simulations.

\begin{figure}[!ht]
\begin{python}
# Declare circuit layout and parameters bounds
c_specs        = {`dimension': 6, `low': 1.,  `high': 100.,  `keep_prob': 0.5}
j_specs        = {`dimension': 6, `low': 99., `high': 1982., `keep_num':  3}
l_specs        = {`dimension': 6, `low': 75., `high': 300.,  `keep_prob': 0.5}
phiOffs_specs  = {`dimension': 4, `values': [0.0, 0.5]}
circuit_params = {`c_specs': c_specs, `j_specs': j_specs, `l_specs': l_specs,
                  `phiOffs_specs': phiOffs_specs}

# Declare circuit searcher settings
general_params = {`solver': 'JJcircuitSimV3', `phiExt': None, `target_spectrum': None}

# Initialize circuit searcher
$\color{deepred}\text{\ttb circuit\_searcher}$ = CircuitSearcher(circuit_params, general_params,
                                   database_path = `Experiments')
\end{python}
\caption{Initialization of the circuit searcher for the 4-local coupler discovery workflow. A circuit with three nodes and an additional ground node has six places for circuit components and can have up to four inductive loops. The circuit component and external flux settings are adjusted accordingly. To ensure symmetry in the ground state spectrum, external fluxes are restricted to the values $0.0\,\Phi_0$ and $0.5\,\Phi_0$. In contrast to the 2-node solver, the external flux settings are stored in \texttt{circuit\_params} rather than \texttt{general\_params}. As the ideal solution is not known, a target spectrum is not provided and the circuit merit will be calculated directly from the shape of the candidate circuit spectrum.}
\label{fig:initialization_fourcoupler}
\end{figure}

We now add a random search task to the circuit searcher (see \figref{fig:add_random_search_task_fourcoupler}).
Since it is rare to find a circuit with double-well ground state spectrum, a large number of 15,000 circuits is sampled.
If such a circuit is found, the merit is calculated as described as in Sec.~\ref{sec:fourcoupler}.
The merit described there is implemented as a separate function and the hyperparameters are made accessible with the \texttt{merit\_options} key of the \texttt{add\_task} method.
A constant offset of 100 is applied to the merit for robustness of the subsequent swarm optimization.
It is later subtracted in the analysis such that the maximum merit is zero.
We note that the generic zero merit is assigned to all circuits that have a small or lacking double-well spectrum, or that fail to evaluate.

\begin{figure}[!ht]
\begin{python}
# Specify settings for the random designer
designer_options = {`max_iters': 15000, `max_concurrent': 65, `batch_size': 100}
    
# Specify merit function
merit_options = {'max_peak': 1.5, 'max_split': 10, 'norm_p': 4, 'flux_sens': True,
                 'max_merit': 100}
    
# Add task to the task queue
computing_task_1 = $\color{deepred}\text{\ttb circuit\_searcher}$.add_task(name = `random_search', 
                         designer = `random',     designer_options = designer_options, 
                         merit    = `DoubleWell', merit_options    = merit_options)
\end{python}
\caption{Random search task for 4-local coupler discovery. This step samples 15,000 circuits and is set up to run on a computing cluster node with 64 cores. The double-well merit function is defined in \textsl{CircuitQuantifier/critic\_double\_well.py} (see GitHub repository [S6]) and its hyperparameters are set in \texttt{merit\_options}.}
\label{fig:add_random_search_task_fourcoupler}
\end{figure}

The filtering step keeps the best two sampled circuits (see \figref{fig:filtering_task_fourcoupler}).
Empirically, this approximately corresponds to the number of double-well circuits that are found in each sampling step.
If less than two sampled circuits have a non-zero merit, one of the circuits without a double-well is kept for subsequent optimization in order to maximize usage of the available resources.
As evident in the constant zero-trajectory at the top of \figref{fig:CouplerSearch}(b), however, a swarm optimization that does not start with a double-well circuit will also not find one.
This observation further validates our approach to start the coupler search with massive, parallelized sampling.

\begin{figure}[!ht]
\begin{python}
# Filter the randomly sampled circuits
filtering_task_1 = circuit_searcher.add_task(name = `filtering', designer = `filter_db',
                                             designer_options = {`num_circuits': 2})
\end{python}
\caption{Declaration of a filtering task to determine the best-performing circuits obtained from the random search task. Out of the 15,000 sampled circuits, two are kept for the next design step.}
\label{fig:filtering_task_fourcoupler}
\end{figure}

The swarm optimization step is defined in \figref{fig:swarm_design_fourcoupler}.
It closely follows the computing step definitions discussed previously.
The execution of the workflow and readout of the database is the same as in the flux qubit example in Sec.~\ref{sec:multi_step_workflows}.1.

\begin{figure}[!ht]
\begin{python}
# Specify settings for the swarm designer
designer_options = {`max_iters': 200, `max_concurrent': 65, `n_particles': 10}
    
# Add task to the task queue
computing_task_2 = $\color{deepred}\text{\ttb circuit\_searcher}$.add_task(name = `swarm_optimization', 
                         designer = `particle_swarms', designer_options = designer_options, 
                         merit    = `DoubleWell',      merit_options    = merit_options,
                         use_library = True)
\end{python}
\caption{Implementation of the swarm optimization step. The declaration of this design step closely follows the L-BFGS-B optimization step in the flux qubit example (see Fig.~\ref{fig:lbfgs_design}). The setting \texttt{n\_particles} specifies the number of particles used in the swarm optimization and corresponds to the number of points in parameter space that are evaluated in parallel per iteration step. The \texttt{merit\_options} for the double-well merit are the same as in the random search step of the 4-coupler search (see \figref{fig:add_random_search_task_fourcoupler}).}
\label{fig:swarm_design_fourcoupler}
\end{figure}


\section{Hamiltonian of 2-node benchmark circuits without linear inductors}
\label{sec:supp:Htwonode}

In the automated discovery benchmark discussed in Sec.~\ref{sec:benchmark}, we consider the set of all two-node circuits with three Josephson junctions and no linear inductors.
For this case, it is straightforward to write out the circuit Hamiltonian:
\begin{equation}
\begin{split}
\label{eq:H_2node}
H_\text{2-node} = & \frac{1}{2} \left(C_n^{-1}\right)_{11} Q_1^2 + \frac{1}{2} \left(C_n^{-1}\right)_{22} Q_2^2 + \left(C_n^{-1}\right)_{12} Q_1 Q_2 - E_{\text{J}11}\cos\left(2\pi \frac{\Phi_1}{\Phi_0}\right) - E_{\text{J}22}\cos\left(2\pi \frac{\Phi_2}{\Phi_0}\right) \\ &-E_{\text{J}12}\cos\left(2\pi \frac{\Phi_1-\Phi_2+\Phi_\text{var}}{\Phi_0}\right),
\end{split}
\end{equation}
where $Q_i, \, i \in \{1,2\},$ are the node charge operators, $\Phi_i, \, i \in \{1,2\},$ are the node flux operators, $\Phi_\text{var}$ is the external flux, and $\Phi_0$ is the flux quantum.
The capacitance matrix $C_n$ has the form
\begin{equation*}
C_n =
\begin{pmatrix}
C_{\Sigma 11} + C_{\Sigma 12} & -C_{\Sigma 12} \\
-C_{\Sigma 12} & C_{\Sigma 22} + C_{\Sigma 12}
\end{pmatrix},
\end{equation*}
where $C_{\Sigma ij}$ is the sum of the intrinsic junction capacitance and the shunt capacitance $C_{ij}$ between nodes $i$ and $j$.

The cosine terms in \eqref{eq:H_2node} can be written as displacement operators in the charge basis.
In this way, the circuit Hamiltonian is computed and solved directly in the charge basis by numerical diagonalization.
For larger and more general circuits, we use the general circuit simulation method referenced in Sec.~\ref{sec:methodology} in the main text.
In particular, that approach is necessary for the 4-local coupler discovery, which requires quantization of three-node circuits with linear inductors.


\section{4-local coupler circuit details for selected search results}
\label{sec:supp:fourcoup_spectra_diagrams}

\begin{figure}[hb]
\includegraphics[width = 1.0\columnwidth]{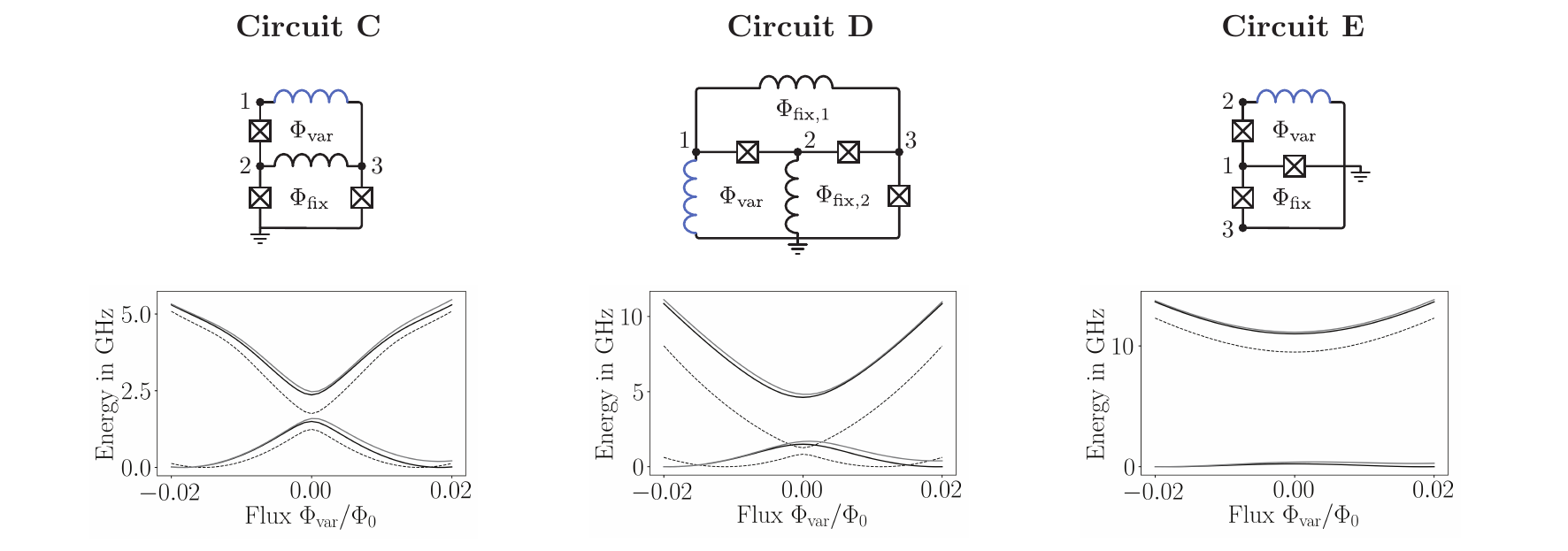}
\caption{
Circuit diagrams and spectra for circuits C, D, E. Top: Network of the inductive circuit components. The inductor that needs to be coupled to four flux qubits to generate 4-local coupling between them is indicated in blue. Circuit parameters are provided in Table~\ref{tab:circuitCDE_params}. Bottom: Ground and first excited state before refinement (dotted black), after refinement (solid black), and with flux offset in a non-primary loop (solid gray).
}
\label{fig:circuitsCDE}
\end{figure}

\begin{figure}[htb]
\includegraphics[width = 1.0\columnwidth]{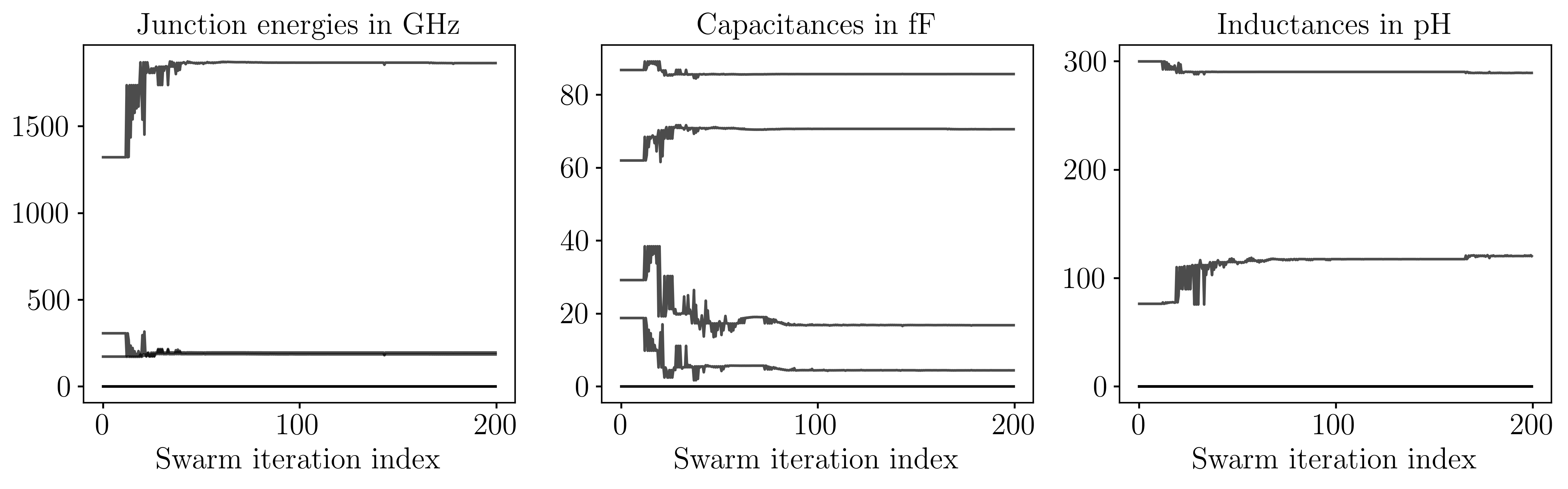}
\caption{
Evolution of the circuit C parameters versus swarm iteration step. At the beginning of the optimization, the starting points of the swarm optimization are generated by applying a small, random offset to the sampled parameter values.
}
\label{fig:parameter_evolution}
\end{figure}

In the discussion of the 4-local coupler search results in Sec.~\ref{sec:fourcoupler}, three circuits are highlighted: circuits C, D, E.
Here we provide additional details about the layout, parameters, and spectra of these circuits.
The inductive network of the circuits and their spectra before refinement, after refinement, and with flux offset in a non-primary loop are shown in \figref{fig:circuitsCDE}.
The circuit parameters before and after refinement are listed in Table~\ref{tab:circuitCDE_params}, and the evolution of the parameters throughout the swarm optimization is shown in \figref{fig:parameter_evolution} for circuit C.
The double-well properties and 4-local coupling strengths (only for refined circuits) are shown in Table~\ref{tab:circuitCDE_properties}.
The 4-local coupling strength is determined in a full system simulation of four qubits and the coupler, which is described in more detail in Sec.~\ref{sec:supp:fullsystem}.

We observe that the swarm optimization step increases the double-well peak and excited state splitting while keeping the asymmetry induced by the non-primary flux offset small.
Therefore, the decreasing (more negative) merit value reflects an improvement of the circuit properties as desired.
The reduced or absent 4-local coupling in the full system simulation of circuits D and E shows that the isolated circuit properties do not translate one-to-one to the coupled circuits.
However, in the case of circuit C, \name was still able to design a coupler with several hundred MHz of 4-local coupling strength using the double-well merit.

\begin{table}[h]
 \centering
 \renewcommand{\arraystretch}{1.2}
 \begin{tabular}{ p{2.2cm} p{1.5cm} p{1.5cm} p{1.5cm} p{1.5cm} p{1.5cm} p{1.5cm} p{1.5cm} p{1.5cm} p{1.5cm} }
    \multicolumn{10}{c}{\textbf{Circuit C}} \\
    \specialrule{.1em}{.05em}{.05em} 
    Parameter& $C_{13}$ & $C_{22}$ & $C_{23}$ & $C_{33}$ & $E_{\text{J}12}/h$ & $E_{\text{J}22}/h$ & $E_{\text{J}33}/h$ & $L_{13}$ & $L_{23}$ \\
    \hline
    Before refine & 98.8\,fF & 15.0\,fF & 38.7\,fF & 67.5\,fF & 1865\,GHz & 187\,GHz & 192\,GHz & 293\,pH & 98.1\,pH \\
    After refine  & 85.7\,fF & 4.46\,fF & 16.8\,fF & 70.6\,fF & 1865\,GHz & 196\,GHz & 185\,GHz & 289\,pH & 120\,pH \\
    \specialrule{.1em}{.05em}{.05em}
    \multicolumn{10}{c}{ } \\
    \multicolumn{10}{c}{\textbf{Circuit D}} \\
    \specialrule{.1em}{.05em}{.05em}
    Parameter& $C_{13}$ & $C_{22}$ & $C_{33}$ & $E_{\text{J}12}/h$ & $E_{\text{J}23}/h$ & $E_{\text{J}33}/h$ & $L_{11}$ & $L_{13}$ & $L_{22}$ \\
    \hline
    Before refine & 86.7\,fF & 5.74\,fF & 77.1\,fF & 683\,GHz & 1005\,GHz & 1039\,GHz & 189\,pH & 158\,pH & 188\,pH \\
    After refine  & 81.5\,fF & 2.66\,fF & 67.0\,fF & 544\,GHz & 902\,GHz & 911\,GHz & 158\,pH & 170\,pH & 166\,pH \\
    \specialrule{.1em}{.05em}{.05em} 
    \multicolumn{10}{c}{ } \\
    \multicolumn{10}{c}{\textbf{Circuit E}} \\
    \specialrule{.1em}{.05em}{.05em} 
    Parameter& $C_{12}$ & $C_{13}$ & $C_{22}$ & $E_{\text{J}11}/h$ & $E_{\text{J}12}/h$ & $E_{\text{J}13}/h$ & $L_{22}$ & $L_{33}$ & \\
    \hline
    Before refine & 66.9\,fF & 42.6\,fF & 68.5\,fF & 1950\,GHz & 767\,GHz & 611\,GHz & 92.7\,pH & 204\,pH & \\
    After refine  & 60.3\,fF & 35.6\,fF & 73.8\,fF & 1905\,GHz & 801\,GHz & 640\,GHz & 85.7\,pH & 188\,pH & \\
    \specialrule{.1em}{.05em}{.05em} 
 \end{tabular}
 \caption{Parameters of circuits C, D, E before and after swarm optimization (refinement).}
 \label{tab:circuitCDE_params}
\end{table}

\begin{table}[h]
 \centering
 \renewcommand{\arraystretch}{1.3}
 \begin{tabular}{p{3.7cm} p{1.8cm} p{1.8cm} p{1.8cm} p{1.8cm}}
    \specialrule{.1em}{.05em}{.05em}
    Circuit instance& $h_\text{peak}$ & $h_\text{split}$ & $h_\text{sens}$ & $2M/2\pi$ \\
    \hline
    Circuit C before refine & 1.24\,GHz & 0.51\,GHz & 0.20\,GHz & \\
    Circuit C after refine  & 1.50\,GHz & 0.87\,GHz & 0.20\,GHz & 0.573\,GHz \\
    \hline
    Circuit D before refine & 0.84\,GHz & 0.42\,GHz & 0.73\,GHz & \\
    Circuit D after refine  & 1.50\,GHz & 3.12\,GHz & 0.67\,GHz & small \\
    \hline
    Circuit E before refine & 0.26\,GHz & 9.23\,GHz & 0.27\,GHz & \\
    Circuit E after refine  & 0.24\,GHz & 10.8\,GHz & 0.27\,GHz & small \\
    \specialrule{.1em}{.05em}{.05em}
 \end{tabular}
 \caption{Double-well properties and 4-local coupling strengths of circuits C, D, E before and after swarm optimization. Truncation of the Hilbert space is estimated to lead to a 1\%-level error in the properties of circuits C, D and a 20\%-level error in the properties of circuit E. The 4-local coupling strength is determined in a full system simulation including four flux qubits and the coupler. A ``small" coupling denotes that no good setting could be found for the coupler-qubit mutual inductances such that the 4-local coupling strength is larger than spurious couplings of other locality.}
 \label{tab:circuitCDE_properties}
\end{table}


\section{Derivation of the circuit C Hamiltonian}
\label{sec:supp:fourcoup_analytic}

In this section, we derive the Hamiltonian of circuit C using the general procedure referenced in Sec.~\ref{sec:methodology} of the main text.
We note that in the automated discovery software, the circuit Hamiltonian is constructed and solved automatically.
By re-deriving it here and numerically comparing the spectrum to the result from \namecomma we show the validity of the automated discovery result and rule out that the software has exploited special cases or bugs in the circuit simulator code that produce unphysical double-well spectra.
In fact, one limitation of earlier versions of the code was that promising search results were shown to be unphysical upon closer examination.
Additionally, the derivation here demonstrates a practical example of the circuit quantization method.

\begin{figure}[ht]
\includegraphics[width = 1.0\columnwidth]{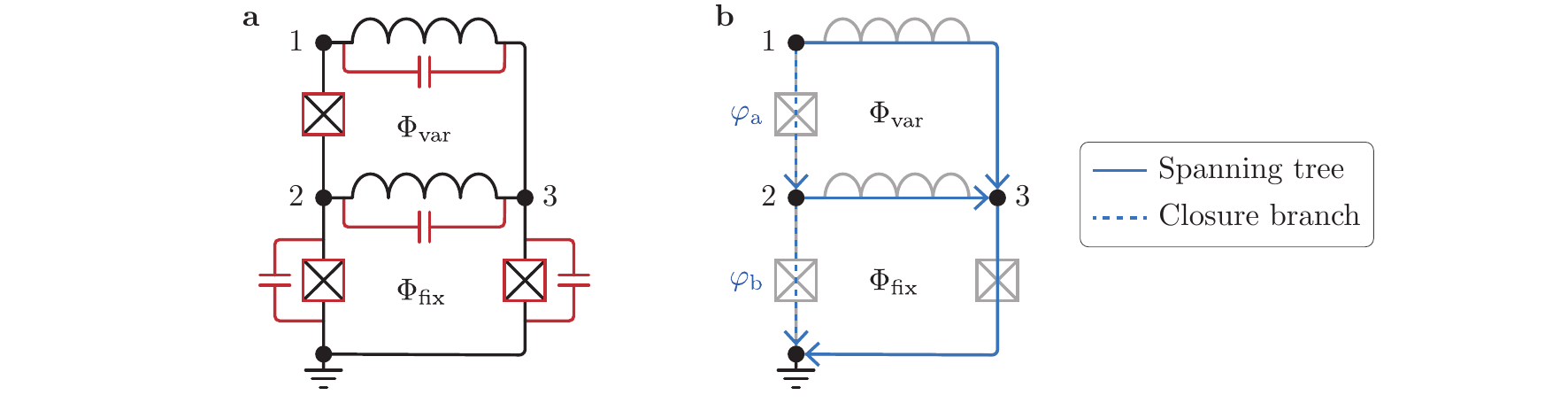}
\caption{
Circuit C network. (a) Circuit diagram including capacitative network, which is highlighted in red. Highlighting the boxes around the junctions emphasizes that the respective junction capacitance is part of the capacitative network. (b) Spanning tree (blue, solid) and closure branches (blue, dashed) of the inductive network.
}
\label{fig:circuitC}
\end{figure}

As described in [S7], the circuit Hamiltonian can be written as the sum of a kinetic (capacitative) term and an inductive potential:
\begin{equation*}
    H = \frac{1}{2}\, \mathbf{Q}^\text{T} \, \text{C}_n^{-1} \, \mathbf{Q} + \frac{1}{2}\, \mathbf{\Phi}^\text{T} \, \text{L}_n^{+} \, \mathbf{\Phi} - \sum_{i,j \in \mathcal{S}_J} E_{\text{J}_{ij}} \cos\left(\varphi_{ij} \left( \mathbf{\Phi} \right) \right),
\end{equation*}
where $\mathbf{Q}$ is the vector of node charges and $\mathbf{\Phi}$ is the vector of node fluxes.
The node capacitance matrix $\text{C}_n$ is constructed from the capacitive network of the circuit, which is shown in red in \figref{fig:circuitC}(a): Diagonal elements are the sum of capacitances connected to a node, and off-diagonal elements are the negative capacitance between two nodes.
We determine the capacitance matrix as
\begin{equation*}
	\text{C}_n =
	\begin{pmatrix}
		C_\text{J12} + C_{13} & -C_\text{J12}                                      & -C_{13} \\
		-C_\text{J12}               & C_{\Sigma22}+C_\text{J12}+C_{23} & -C_{23} \\
		-C_{13}                       & -C_{23}                                              & C_{\Sigma33}+C_{13}+C_{23}
	\end{pmatrix},
\end{equation*}
where we defined $C_{\Sigma ij} = C_{ij} + C_{\text{J}ij}$.
The pseudo-inverse node inductance matrix is determined by placing the inverses of all linear inductances connected to a node on the diagonal and the negative inverse inductance between two nodes on the off-diagonal:
\begin{equation*}
	\text{L}_n^+ =
	\begin{pmatrix}
		\frac{1}{L_{13}}  & 0                        & -\frac{1}{L_{13}} \\
		0                        & \frac{1}{L_{23}}  & -\frac{1}{L_{23}} \\
		-\frac{1}{L_{13}} & -\frac{1}{L_{23}} & \frac{1}{L_{13}} + \frac{1}{L_{23}}
	\end{pmatrix}.
\end{equation*}
The set $\mathcal{S}_J$ contains the pairs of indices for all nodes that have a junction between them and $\varphi_{ij}\left(\mathbf{\Phi}\right)$ denotes the phase difference across the junction between nodes $i$ and $j$.

In the next step, we impose the boundary conditions of fluxoid quantization.
This requires the definition of a spanning tree and closure branches.
As all nodes are connected inductively, the spanning tree and closure branch definitions are only necessary for the circuit's inductive network.
We choose their configuration as shown in \figref{fig:circuitC}(b).
The closure branches are chosen across junctions J$_{12}$ and J$_{22}$ and we label their respective phase differences as $\varphi_a$ and $\varphi_b$.
Fluxoid quantization yields the following relations:
\begin{align*}
	\varphi_a &= \phi_1 + \phi_2 - 2\phi_3 - \phi_\text{var}, \\
	\varphi_b &= \phi_2 - \phi_\text{fix}.
\end{align*}

Here we defined $\phi_\lambda = 2\pi\frac{\Phi_\lambda}{\Phi_0}, \, \lambda \in \{1,2,3,\text{var},\text{fix}\}$.
The inductance matrix has rank 2, so the canonical variables can be transformed to write $\text{L}_n^+$ as a single $2\times2$ block.
This is achieved with the following transformation matrix:
\begin{equation*}
	\text{R} =
	\begin{pmatrix}
		1 & 0 & 1 \\
		0 & 1 & 1 \\
		1 & 1 & 1
	\end{pmatrix}.
\end{equation*}
The variables and matrices are transformed as follows:
\begin{align*}
	\tilde{\mathbf{\Phi}}                  &= \text{R}^{-1} \, \mathbf{\Phi}, \\
	\tilde{\mathbf{Q}}                     &= \text{R} \, \mathbf{Q}, \\
	\tilde{\text{L}}_n^+     &= \text{R} \, \text{L}_n^+ \, \text{R}, \\
	\tilde{\text{C}}_n^{-1} &= \text{R}^{-1} \, \text{C}_n^{-1} \, \text{R}^{-1}.
\end{align*}
The matrix $\tilde{\text{L}}_n^+$ is diagonal with only two non-zero entries, which are the inductances of the oscillator modes of the circuit:
\begin{equation*}
	\tilde{\text{L}}_n^+ =
	\begin{pmatrix}
		\frac{1}{L_{23}} & 0 & 0 \\
		0 & \frac{1}{L_{13}} & 0 \\
		0 & 0 & 0
	\end{pmatrix}.
\end{equation*}
Therefore, we write the transformed nodes $\tilde{1}$ and $\tilde{2}$ in the oscillator basis and nodes $\tilde{3}$, which is not connected to a linear inductance, in the charge basis.
The frequencies and impedances of the oscillator modes are calculated as follows:
\begin{align*}
	\omega_1 &= \sqrt{ \frac{ \tilde{c}_{11}^{-1} }{ L_{23} } }, \,\, Z_1 = \sqrt{ L_{23} \tilde{c}_{11}^{-1} }, \\
	\omega_2 &= \sqrt{ \frac{ \tilde{c}_{22}^{-1} }{ L_{13} } }, \,\, Z_2 = \sqrt{ L_{13} \tilde{c}_{22}^{-1} },
\end{align*}
where $\tilde{c}_{ij}^{-1}$ denotes the entries of the rotated inverse capacitance matrix $\tilde{\text{C}}_n^{-1}$.

We also apply the variable transformation to the flux variables in the Josephson potential terms.
The cosines are then written as exponentials, which are recognized as quantum optical displacement operators $D_i\left(\alpha_i\right)$ for the nodes written in the oscilator basis:
\begin{equation*}
\begin{split}
    \cos\left( 2\pi \frac{\Phi_i}{\Phi_0} \right) = &\,\frac{1}{2} \left( D_i^\dag\left(\alpha_i\right) + D_i\left(\alpha_i\right) \right), \, i \in \{\tilde{1}, \tilde{2}\}, \\
         &\,\alpha_i = i\frac{2\pi}{\Phi_0}\sqrt{\frac{\hbar Z_i}{2}}.
\end{split}
\end{equation*}
In the charge basis, the exponentials transform to charge displacement operators $d_3$ for node $\tilde{3}$:
\begin{equation}
\label{eq:charge_displacement}
    \cos\left( 2\pi \frac{\Phi_3}{\Phi_0} \right) = \frac{1}{2} \left( d_3^\dag + d_3 \right).
\end{equation}
Thus, all the inductive terms in the circuit Hamiltonian are written in the desired mixed-basis representation of oscillator and charge basis.
We are left with the capacitative terms that include off-diagonal elements of $\tilde{\text{C}}_n^{-1}$.
These are of the form $\cel{ij} \tilde{Q}_i \tilde{Q}_j, \, i \neq j$.
The charge operators of nodes $\tilde{1}$ and $\tilde{2}$ are written in the oscillator basis, and the charge operator of node $\tilde{3}$ is kept and represented in the charge basis.

The circuit Hamiltonian after transformation and expression in the mixed-basis representation separates into terms of the oscillator modes $\tilde{1}$ \& $\tilde{2}$ only, charge mode $\tilde{3}$ only, and interactions between the oscillator and charge modes.
To structure the lengthy Hamiltonian expression, we separate it into a sum of an oscillator Hamiltonian $H_\text{O}$, charge Hamiltonian $H_\text{C}$, and interaction Hamiltonian $H_\text{int}$:
\begin{align*}
	H                = & \, H_\text{O} + H_\text{C} + H_\text{int}, \\
	H_\text{O}  = & \, \hbar \omega_1 a_1^\dag a_1 + \hbar \omega_2 a_2^\dag a_2  - \frac{\hbar}{2\sqrt{Z_1 Z_2}} \cel{12} \left( a_1 - a_1^\dag \right) \left( a_2 - a_2^\dag \right) - \frac{E_{\text{J}12}}{2} \left( e^{i\phi_\text{var}}  D_1^\dag\left(\alpha_1\right) D_2^\dag\left(\alpha_2\right) + \text{h.c.} \right),  \\
	H_\text{C}  = & \, \frac{1}{2} \cel{33} \, \tilde{Q}_3^2,  \\
	H_\text{int} = &-i\sqrt{\frac{\hbar}{2Z_1}} \cel{13} \left( a_1 - a_1^\dag \right) \tilde{Q}_3 - i\sqrt{\frac{\hbar}{2Z_2}} \cel{23} \left( a_2 - a_2^\dag \right) \tilde{Q}_3 - \frac{E_{\text{J}22}}{2} \left( e^{-i\phi_\text{fix}} D_2^\dag\left(\alpha_2\right) d_3^- + \text{h.c.} \right), \\
	                     &- \frac{E_{\text{J}33}}{2} \left( D_1^\dag\left(\alpha_1\right) D_2^\dag\left(\alpha_2\right) d_3^- + \text{h.c.} \right). \\
\end{align*}
Here, $a_i^\dag$ and $a_i$ are the bosonic raising and lowering operators and ``h.c." is short for the hermitian conjugate of the preceding term. We simulate the spectrum of this Hamiltonian in Python and confirm that it matches the one calculated by the circuit simulator in \namecomma which is shown in \figref{fig:circuitsCDE}.


\section{Full system simulation of four qubits and coupler}
\label{sec:supp:fullsystem}

In Sec.~\ref{sec:fourcoupler}, a full system simulation is presented in which four flux qubits are mutually inductively coupled to circuit C.
The simulated circuit is shown in \figref{fig:CouplerSearch}(c) and its spectrum for commonly sweeping the external flux of the qubits around degeneracy is shown in \figref{fig:CouplerSearch}(d).
The simulation shows that circuit C mediates a 4-local interaction between the qubits and confirms that the double-well merit can be used as a proxy to optimize such a coupling.
Here we provide additional details on the parameters and simulation technique for the full system simulation.

The qubits are three-junction flux qubits with a linear loop inductance and a shunt capacitance across the small junction.
The small junction has a Josephson energy of $h\times 119\,\text{GHz}$ and the large junction an energy of $h\times 183\,\text{GHz}$.
A linear inductance of $100\,\text{pH}$ and shunt capacitance of $50\,\text{fF}$ are chosen.
With those parameters, the qubits are close to the C-shunt flux qubit regime [S8].
Their persistent current is $226\,\text{nA}$ and the linear inductance is chosen to allow for sufficient mutual inductive coupling to shift the coupler flux between its bias points along the double-well ground state spectrum.

The full system Hamiltonian is determined via the general method referenced in Sec.~\ref{sec:methodology}, with the mutual inductive couplings between the subsystems included in the pseudo-inverse inductance matrix $\text{L}_n^+$.
As the size of the Hamiltonian is too large for a Hilbert space truncation that preserves accuracy of the simulation, a hierarchical diagonalization approach is adopted in which the subsystems are diagonalized first.
Since the subsystems are weakly coupled relative to the intra-subsystem modes, they can be combined to the full system in a second diagonalization step.

The mutual inductance required to match the flux shifts induced by the qubit excitation manifolds on the coupler with the double-well bias points for 4-local coupling (red dots in \figref{fig:CouplerSearch}(a)) is not known a priori.
Therefore, the optimal qubit-coupler mutual inductance is determined in a parameter sweep and found to be $38.25\,\text{pH}$.
We note that the mutual inductance can be lowered at the expense of raising the qubit persistent current, which would increase the qubits' flux noise sensitivity.
Finally, we point out that the design of the full system requires further investigation.
The goal of this work, however, primarily lay in finding a circuit that exhibits the desired spectral properties and holds promise for experimental implementation of a 4-local interaction between qubits.

\bibsection{}

\begin{itemize}
    \item[{[S1]}] J. L. Morales and J. Nocedal, ACM Transactions on Mathematical Software (TOMS) \textbf{38}, 7 (2011).
    \item[{[S2]}] A. S. Lewis and M. L. Overton, Submitted to SIAM Journal on Optimization (2009).
    \item[{[S3]}] Y. Shi and R. Eberhart, in \textit{International Conference on Evolutionary Computation Proceedings. IEEE World Congress on Computational Intelligence} (IEEE, 1998) pp. 69--73.
    \item[{[S4]}] R. Eberhart and J. Kennedy, in \textit{Proceedings of the Sixth International Symposium on Micro Machine and Human Science} (IEEE, 1995) pp. 39--43.
    \item[{[S5]}] L. J. V. Miranda et al., Journal of Open Source Software \textbf{3}, 21 (2018).
    \item[{[S6]}] The automated discovery implementation for use with a generic circuit simulator (not provided) is available at the following address: \url{https://github.com/aspuru-guzik-group/scilla}.
    \item[{[S7]}] A. J. Kerman, \textit{Efficient, hierarchical simulation of complex Josephson quantum circuits,} in preparation.
    \item[{[S8]}] F. Yan, S. Gustavsson, A. Kamal, J. Birenbaum, A. P. Sears, D. Hover, T. J. Gudmundsen, D. Rosenberg, G. Samach, S. Weber, \textit{et al.}, Nature Communications \textbf{7}, 12964 (2016).
\end{itemize}